\documentclass[fleqn,usenatbib]{mnras}

\usepackage[T1]{fontenc}
\usepackage{ae,aecompl}

\usepackage{graphicx}	
\usepackage{amsmath}	
\usepackage{amssymb}	
\usepackage{anyfontsize}

\newcommand{\Eq}[1]{Equation\,(\ref{#1})}
\newcommand{\Fig}[1]{Fig.~\ref{#1}}

\defcitealias{Youdin2005}{YG05}
\defcitealias{Squire2018}{SH18}
\defcitealias{Krapp2019}{K19}

\title[Dust Settling Instability]{Dust Settling Instability in Protoplanetary Discs}

\author[Krapp, L., Youdin, A. N., Kratter, K. M., Ben\'itez-Llambay, P.]{\fontsize{13}{1}\selectfont{
Leonardo Krapp,$^{1}$\thanks{E-mail: krapp@email.arizona.edu}
Andrew N. Youdin,$^{1,2}$
Kaitlin M. Kratter,$^{1}$ and
Pablo Ben\'itez-Llambay$^{3}$}
\\
$^{1}$Department of Astronomy and Steward Observatory, University of Arizona, Tucson, Arizona 85721\\
$^{2}$The Lunar and Planetary Laboratory, University of Arizona\\
$^{3}$Niels Bohr International Academy, Niels Bohr Institute,
Blegdamsvej 17, DK-2100 Copenhagen \O{}, Denmark
}

\date{Accepted 2020 June 19. Received 2020 June 18; in original form 2020 April 4}

\pubyear{2020}

\hypersetup{draft}

\begin{document}
\label{firstpage}
\pagerange{\pageref{firstpage}--\pageref{lastpage}}
\maketitle

\begin{abstract}
The streaming instability (SI) has been extensively studied in the linear and non-linear regimes as a mechanism to concentrate solids and trigger planetesimal formation in the midplane of protoplanetary discs. A related dust settling instability (DSI) applies to particles while settling towards the midplane. The DSI has previously been studied in the linear regime, with predictions that it could trigger particle clumping away from the midplane. 
This work presents a range of linear calculations and non-linear simulations, performed with FARGO3D, to assess conditions for DSI growth. We expand on previous linear analyses by including particle size distributions and performing a detailed study of the amount of background turbulence needed to stabilize the DSI. When including binned size distributions, the DSI often produces converged growth rates with fewer bins than the standard SI. With background turbulence, we find that the most favorable conditions for DSI growth are weak turbulence, characterized by $\alpha \lesssim 10^{-6}$ with intermediate-sized grains that settle from one gas scale height.  These conditions could arise during a sudden decrease in disc turbulence following an accretion outburst.
Ignoring background turbulence, we performed a parameter survey of local 2D DSI simulations.  Particle clumping was either weak or occurred slower than particles settle.  Clumping was reduced by a factor of two in a comparison 3D simulation. Overall, our results strongly disfavor the hypothesis that the DSI significantly promotes planetesimal formation.  Non-linear simulations of the DSI with different numerical methods could support or challenge these findings.
\end{abstract}




\section{Introduction}

The behavior of solids in young, protoplanetary discs (PPDs) is of paramount importance for interpreting observations and developing theories of planet formation. 
The settling of  solids -- dust and ice grains and larger pebble-sized agglomerates -- towards the midplane is likely one of the first  steps in the assembly of planetesimals and planetary cores. 
Thus many previous works have focused on dust dynamics near the mid-plane of PPDs, where dust densities are naturally highest and could trigger a range of instabilities, including gravitational and vertically shearing 
\citep{saf69, Goldreich1973,WEIDENSCHILLING1984, Goodman2000, Youdin2002, Gomez2005}. 
In particular, the streaming instability (SI) has been proposed as a mechanism to concentrate solids and trigger  gravitational collapse into planetesimals \citep[][hereafter \citetalias{Youdin2005}]{Youdin2005}.

However, PPD mid-planes are not the only location where dust-gas instabilities might occur.  \citet{Squire2018letter} showed that dust moving through gas can trigger a variety of Resonant Drag Instabilities (RDIs), whenever the frequency of a fluid wave matches the pattern speed of dust drifting along that wave.   While RDI theory strictly holds in the limit of small dust-to-gas ratios, it can usually be extrapolated to larger dust-to-gas ratios and thus larger growth rates.

RDI theory reproduces the SI and also predicts a related dust settling instability (hereafter DSI) \citep[][hereafter \citetalias{Squire2018}]{Squire2018}.  The DSI includes both the radial drift of the SI and the vertical settling of grains, which becomes significant far from the disk midplane.  The DSI also operates without radial drift \citep{Zhuravlev2019}, e.g.\ in a disc radial pressure maximum \citep{Pinilla2017}.  
In the absence of rotation (Keplerian for the DSI) particle sedimentation does not produce a linear instability, but can trigger a non-linear drafting instability that clumps solids \citep{Lambrechts2016}.

\citetalias{Squire2018} predicted that the DSI could concentrate solids away from the midplane, based on rapid linear growth rates at one gas scale height.    
\citetalias{Squire2018} further suggest that the DSI might kickstart planetesimal formation either by triggering gravitational collapse directly or by creating particle concentrations that settle to the midplane and seed the SI.
Even without significant particle concentration, the DSI might trigger weak turbulence away from the midplane, which could affect the coagulation of dust grains as they settle  \cite[e.g.,][]{Krijt2016, Blum2018}.   
Dust growth away from the midplane also affects opacities that are important for PPD observations  \citep[e.g.,][]{DAlessio2006, Rettig2006}.

While the linear growth properties of the DSI are intriguing, non-linear simulations are needed to determine the expected levels of particle clumping and turbulence.  \citetalias{Squire2018} argue that previous simulations of gas and dust in PPDs lacked the resolution and or vertical extent to capture the DSI, motivating the targeted simulations presented here.
Moreover, since DSI growth is strongest on the smallest scales, it is crucial to examine the stabilizing effects of turbulence.  
While the effects of turbulence on the linear DSI have been studied in various limiting cases (\citetalias{Squire2018}, \citealp{Zhuravlev2020}), a systematic exploration of turbulent stability limits is needed and presented in this work.  
Finally, since dust grains have a range of sizes, and particle size distributions significantly influence SI growth rates  \cite[][hereafter \citetalias{Krapp2019}]{Krapp2019}, we examine how particle size distributions affect the DSI.

This work is organized as follows. In Section\,\ref{sec:Equations} we present the basic analytic model of the DSI. 
We extend the work of \citetalias{Squire2018} by exploring the linear phase of the DSI for particle-size distributions rather than monodisperse populations in Section\,\ref{sec:linear}. 
In Section\,\ref{sec:NonLinear}, we analyze the non-linear growth and saturation of the instability using 2D and 3D numerical simulations.
We then reconsider the behavior of the DSI in the presence of a turbulent background in Section\,\ref{sec:Turbulent_Background}, identifying a very limited parameter space in which the linear phase of the instability persists at all. We apply our results to planetesimal formation and dust evolution in Section\,\ref{sec:Discussion}. Future perspectives and major caveats of our work are addressed in Section\,\ref{sec:future}.

\section{Model Equations and Control Parameters}
\label{sec:Equations}

In this section, we present the model equations that reproduce the DSI of \citetalias{Squire2018}.  
We generalize their model of a gas and dust disc to include $N$ dust-species, following the framework of  \cite{Benitez-Llambay2019} (see their Section 3.5).  
This local model is centered at a height $z_0$ below\footnote{Without loss of generality, below is chosen so settling speeds are positive.} the midplane to include vertical particle sedimentation, the crucial ingredient that distinguishes the DSI of \citetalias{Squire2018} from the SI of \citetalias{Youdin2005}.

To introduce the problem,  we present the relevant scales and dimensionless control parameters.  
Like the SI, the DSI relies on the pressure support of the gas disc.  In hydrostatic equilibrium, radial pressure gradients, which are negative in an unperturbed disc, reduce the orbital speed of the gas below the Keplerian speed, $v_K$.  
The amount of this reduction,  $\eta v_K$, sets the velocity scale of the problem, where $\eta$ is one-half of the ratio between the radial pressure  and  gravitational forces \citep[see e.g.,][]{Nakagawa1986}.

The effects of compressibility are parameterized\footnote{In other works \citep[e.g.,][]{Bai2010ab,Abod2019} where $H$ is the scale length, $\Pi$ sets the strength of the radial pressure support.  Since we normalize to the radial pressure gradient, our interpretation in terms of compressibility is more appropriate.}   by $\Pi = \eta v_K/c_{\rm s}$, where $c_{\rm s} = H \varOmega_0$ is the sound speed, $H$ is the gas scale height and $\varOmega_0 = v_K/r_0$, the orbital frequency at the reference disc radius $r_0$. 
The timescale is given in units of $\varOmega^{-1}_0$, and because the velocity scale is $\eta v_K$, the reference lengthscale is thus $\eta r_0$, which is ($3/2$ times) the radial distance between Keplerian and pressure supported orbits of the same speed.
\citetalias{Squire2018} assumed $\eta = (H/r_0)^2$, fixing $\eta = 10^{-3}$ in the numerical evaluations.  
We adopt this value in our numerical examples to facilitate comparison. 
Thus, the corresponding numerical value of our fiducial parameter is $\Pi = 10^{-3/2}$.  

We define the dimensionless reference height for the dust particles settling via the parameter $\zeta = z_0/(\eta r_0)$.  
In well-mixed gas-dust regimes, particles settle from $z_0 \lesssim H$ if turbulent diffusion is strong enough to remove grains from the mid-plane \citep[e.g.,][]{Dubrulle1995,YoudinLithwik2007}. 
Settling from $H$ without (or with very weak) turbulence is not a consistent equilibrium state due to particle sedimentation.  We discuss in Section\,\ref{sec:Turbulent_Background} how this initial condition could arise from time-variable accretion and turbulence. 
This work considers the limit of no background turbulence, until Section\,\ref{sec:Turbulent_Background}.

Because the DSI growth-rate increases with the vertical height, consistent with the \citetalias{Squire2018} linear analysis, we fix $z_0 = H$ to obtain the fastest possible growth. 
Note that this assumption gives $\zeta = H/(\eta r_0) = \eta^{-1/2}$.  
Thus many of the analytic scalings from \citetalias{Squire2018} can readily be generalized to different $z_0$ by the replacement $\eta^{1/2} \rightarrow 1/\zeta$.\footnote{This replacement is valid because the dynamics is nearly incompressible and thus $\Pi \rightarrow 0$ should not appear to lowest order.}   

The variables describing gas and dust $j$th-species density and velocity are defined as $\rho_{\rm g}$, ${\bf v}_{\rm g}$, and $\rho_j$, ${\bf v}_{j}$, respectively.
For each dust species indexed by $j = 1, ..., N$, we define the steady-state density $\rho_j^{0}$, and the aerodynamic stopping time as $t_{{\rm stop},j}$.  
This drag time is approximately constant if gas density varies little (appropriate for the local and nearly incompressible motions considered here as is shown in Section\,\ref{sec:NonLinear}) and the drag law is linear in velocity (appropriate for small dust and pebbles, see e.g.\ \citealp{Adachi1976, Chiang2010}).  
Thus, for each of the $N$ dust species, we introduce two dimensionless parameters: $T_{{\rm s}j} = \varOmega_0 t_{{\rm stop},j}$ and  $\epsilon_j \equiv \rho_j^{0}/\rho_{\rm g}^{0}$, with $\rho^0_{\rm g}$ the steady-state gas density  at $z=z_0$.
The drag acceleration on each dust species $j$ and the back-reaction on the gas are

\begin{align}
{\bf F}_{j} &=  -\varOmega_0\frac{1}{T_{{\rm s}j}} \left({\bf v}_j - {\bf v}_{\rm g} \right) \\
{\bf F}_{\rm g} &= \frac{1}{\rho_{\rm g}}\varOmega_0\sum_{k=1}^{N} \frac{\rho_k}{T_{{\rm s}k}} \left( {\bf v}_k - {\bf v}_{\rm g} \right)  \, ,
\end{align}
\citep[e.g.,][]{Epstein1923,Whipple1972}. 
Since the model is vertically local (in addition to being horizontally local), the vertical gravitational acceleration is constant and added to  all the dust-species as
\begin{equation}
{\bf a}_{\rm d} = z_0\varOmega^2_0\mathbf{e}_z\,, 
\label{Eq:accelerations_d}
\end{equation}
with $\mathbf{e}_z$ the vertical unit vector, and acceleration is positive for the region below the midplane.
The gas also experiences an external acceleration of %
\begin{align}
{\bf a}_{\rm g} = 2\eta v_K\varOmega_0  \mathbf{e}_x - \epsilon z_0\varOmega^2_0\mathbf{e}_z\,, \quad \quad 
\label{Eq:accelerations_g}
\end{align}
with $\mathbf{e}_x$ the radial unit vector and $\epsilon = \sum_j \epsilon_j$. 
The radial acceleration term in this equation (familiar from studies of the SI)  must be added to the local model to account for acceleration by a global radial pressure gradient \citep[see e.g.][]{Youdin2007}.  There is no corresponding term for large scale vertical pressure gradients, which hydrostatically balance the vertical gravity, so the terms cancel.
The vertical acceleration above arises instead from the effects of vertical particle settling.  
Each particle species has an equilibrium (terminal) velocity of $v_{{\rm sett}, j} = \varOmega_0 z_0 T_{{\rm s}j}$ which causes a back reaction acceleration of $\epsilon_j \varOmega_0^2 z_0$ on the gas.
The pressure gradient that balances this acceleration in equilibrium must be added by hand to a local model.  
Summing over species gives the vertical term in Eq. \eqref{Eq:accelerations_g}.   Note that there is no corresponding radial  back reaction term, because radial (and azimuthal) drift is not balanced hydrostatically, but by gas flow (i.e. \citealp{Nakagawa1986}).

Treating the dust species as a pressureless fluid (\citetalias{Youdin2005}) in a Keplerian shearing box \citep{Goldreich1965}, the gas and dust continuity and momentum equations are:
\begin{align} 
\partial_t \rho_{j} + \nabla \cdot\left( \rho_{j} \mathbf{v}_{j}\right) &= 0\,, \label{Eq:set_begin} \\
\partial_t \rho_{\rm g} + \nabla \cdot\left( \rho_{\rm g} \mathbf{v}_{\rm g}\right) &= 0\,,\\
\partial_t \mathbf{v}_{j} + \mathbf{v}_{j} \cdot \nabla \mathbf{v}_{j} = & \,\, 3\varOmega_0^2 x \mathbf{e}_x - 2\varOmega_0{\bf e}_z\times \mathbf{v}_{j} + {\bf F}_j + {\bf a}_{\rm d}\,, \\
\partial_t \mathbf{v}_{\rm g} + \mathbf{v}_{\rm g} \cdot \nabla \mathbf{v}_{\rm g}  =& \,\, 3\varOmega_0^2 x \mathbf{e}_x  -  2\varOmega_0 {\bf e}_z \times \mathbf{v}_{\rm g} + {\bf F}_{\rm g} + {\bf a}_{\rm g} \nonumber \\ 
&-  \frac{\nabla P}{\rho_{\rm g}}\,,
\label{Eq:set_end}
\end{align}
for $j=1,\dotsc,N$.
The gas pressure, $P$, is defined as $P=c^2_{\rm s}\rho_{\rm g}$ with constant sound speed in our isothermal analysis.

As we mentioned, this multi-species framework is adopted from \cite{Benitez-Llambay2019}.  
Setting the number of dust species to one and, with changes to the included accelerations, these model equations\footnote{The RHS of the dust momentum equation is also consistent with a Lagrangian super-particle model.} have been used to study the SI \citep{Youdin2007, Johansen&Youdin2007} and a variety of RDIs  \cite{Seligman2019,Moseley2019, Hopkins2019} in the linear and nonlinear regime.
\begin{figure}
	\includegraphics[]{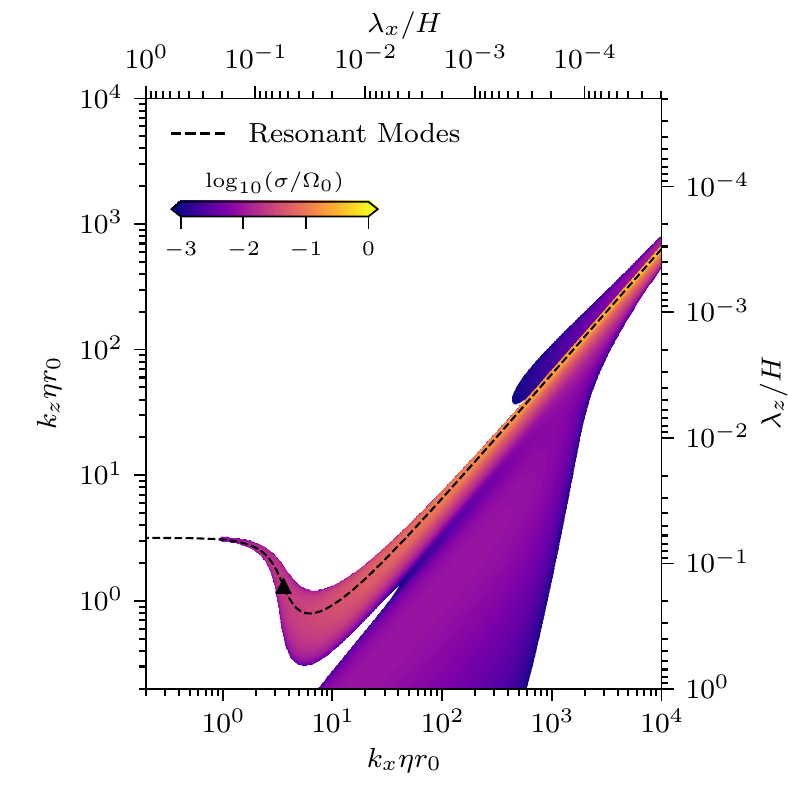}
	\caption{The colormap shows the growth rate, $\sigma$, as a function of the normalized wavenumber (and wavelength) for a dust-species with $T_{\rm s1}=10^{-2}$ and $\epsilon=10^{-3}$. The white regions indicate  stable modes ($\sigma < 0$) or secular modes ($\sigma < 10^{-3}\varOmega_0$).
	The dashed line shows the resonant condition (see Eq.\,\ref{eq:resonances}) where laminar growth rates are largest. 
	The dark triangle demarcates a case considered by SH18 (and utilized for comparison here) with $\theta_k = \tan^{-1}(k_x/k_z) = 70^{\circ}$ and $\sigma(\theta_k)=0.023\varOmega_0$. 
	Due to the absence of stabilizing effects (i.e.\ turbulence) on small scales, growth rates increase without bound as  $k$ increases along the resonant condition. } \label{fig:fig1}
\end{figure}

The local approximation of this model does not include gas or dust vertical stratification. Thus, the applicability is limited to length scales where the vertical gravitational acceleration is nearly constant and to timescales less than the settling time across this length.
In PPDs, such length scales are much smaller than the gas ($H$) or dust scale heights. 
Therefore, we limit our study to vertical wavelengths smaller than $H$, consistent with the adopted shearing-box equations \citep[see e.g.,][]{Latter2017}.
We furthermore limit our analysis to small solids with $T_{\rm s} < 0.5$
so that particles settle gradually towards -- instead of oscillating about -- the mid-plane \citep[e.g.,][]{Weidenschilling1980, Youdin2010}.  
Such oscillations would undermine the fluid approximation for solids. 

\section{Linear stability}

\label{sec:linear}
We use equations \eqref{Eq:set_begin}-\eqref{Eq:set_end} to study the linear growth of the DSI.  Details of the linear analysis are presented in Appendix\,\ref{sec:linearized}.
In Section\,\ref{sec:mono} we reproduce the case of a single dust-species, before generalizing it to multiple dust-species in Section\,\ref{sec:poly}, following the techniques of \citetalias{Krapp2019}.

Our analysis is axisymmetric for simplicity, to facilitate comparison with previous work, and to focus on exponential growth instead of the transient amplification of non-axisymmetric shearing waves.  
Waves in our local model thus have a radial, $k_x = 2\pi/\lambda_x$, and a vertical, $k_z = 2\pi/\lambda_z$ wavenumber, with $k^2 = k^2_x + k^2_z$.  
While the system supports multiple unstable eigenmodes, we present the fastest growing mode, which we identify as a DSI mode because, in the single species case, growth matches the predictions of the RDI theory of \citetalias{Squire2018} (see Section\,\ref{sec:mono}).
In our study we always include the radial external acceleration, however, a pure settling instability occurs in the case with no radial-drift \citep{Zhuravlev2019}. This may be relevant for PPDs at pressure bumps where the analysis presented by \cite{Auffinger2018} is more suitable.

\begin{figure*}
	\includegraphics[]{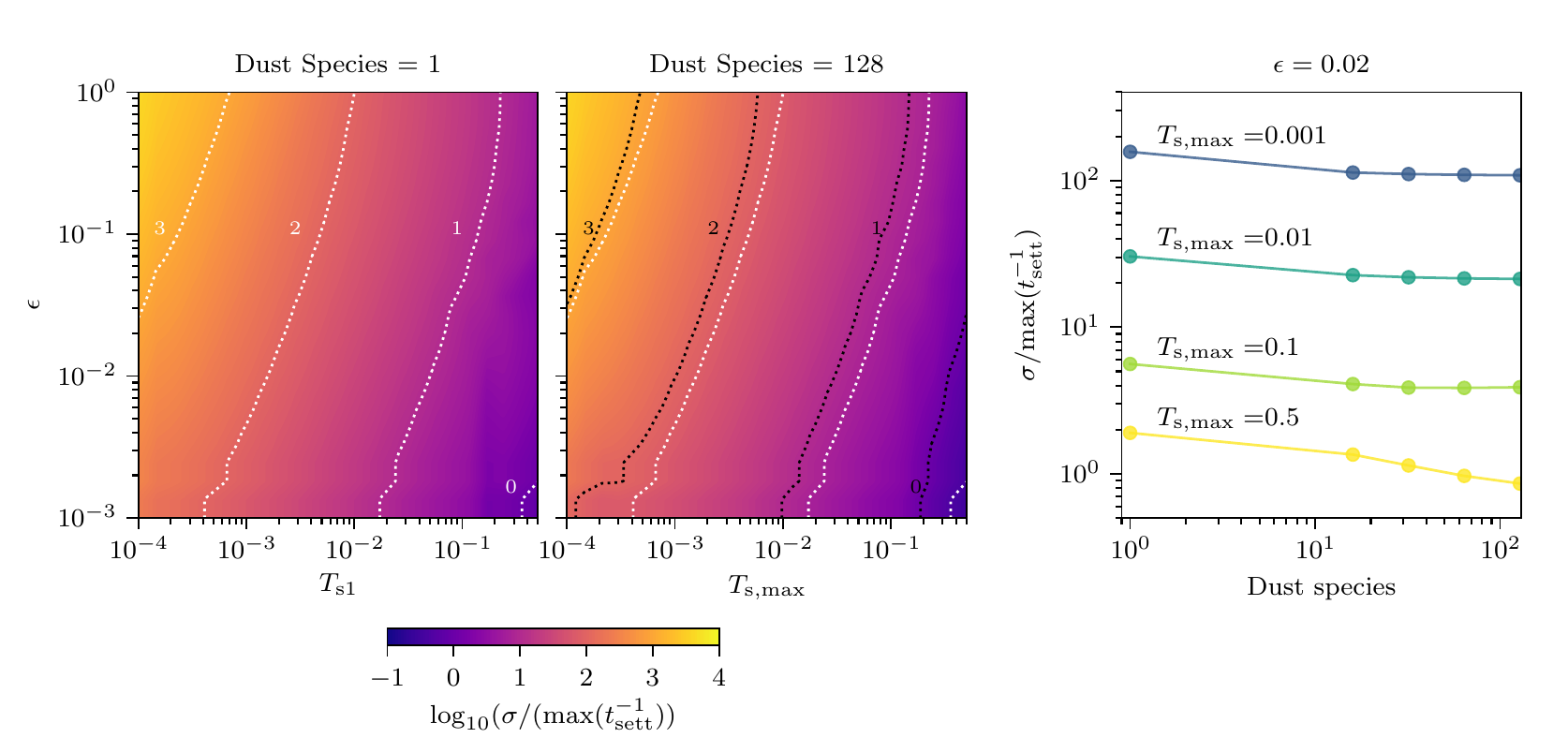}
	\caption{These plots show the effect of a dust size distribution, and of the number of species used to resolve that size distribution, on the maximum growth rate of the DSI, normalized to the settling rate of the largest grains.
		\emph{(Left)} Growth rates for a single dust species vs. dust-to-gas mass ratio, $\epsilon$ and stopping time, $T_{\rm s}$, where contours are labeled by the log of the normalized growth rate.
		\emph{(Center)} Growth rates for a size distribution with 128 species from $T_{\rm s, min} = 10^{-5}$ to the labeled  $T_{\rm s, max}$. 
		Spreading the same total dust-to-gas mass ratio over a size distribution slightly reduces growth rates. 
		The dotted lines demarcate growth rate contours for the single species case (white), and multi-species  case (black). Comparing the two shows that the multi-species case requires a larger $\epsilon$ to reproduce the same growth rate (at fixed maximum particle size).
		\emph{(Right)} Convergence test for growth rate vs.\ number of species for fixed $\epsilon = 0.02$ and several $T_{\rm s, max}$.  Only the largest $T_{\rm s, max} = 0.5$ is not well-converged for 128 species. 
		}	\label{fig:fig2}
\end{figure*}

\subsection{Monodisperse dust populations}\label{sec:mono}

Fig.\,\ref{fig:fig1} shows the DSI growth rates for particles with a Stokes number\footnote{Since, we consider only a single species here, we drop the subscript to label that species.} $T_{\rm s}=10^{-2}$ and dust-to-gas mass ratio $\epsilon = 10^{-3}$.
Particles are assumed to settle from a height $z_0 = H$.
The parameters match a case considered by \citetalias{Squire2018} in their Figure 4 and Equation 32.  

The growth rates in Fig.\,\ref{fig:fig1} are largest along the dashed curve, which follows the resonant condition of \citetalias{Squire2018}:
\begin{equation}
{\bf k} \cdot {\bf w}  = \frac{\pm k_z}{ \sqrt{k^2_x + k^2_z}} \varOmega_0.
\label{eq:resonances}
\end{equation}
This condition matches the Doppler frequency $\bf{k}\cdot \bf{w}$ on the left hand side -- for the drift speed ${\bf w}$ defined in \Eq{eq:eqdrift} -- with the Epicyclic  frequency on the right hand side.  

The dashed curve is obtained after solving the resonant condition (Eq.\,\eqref{eq:resonances}) for  $k_{z}$, at a given $k_x$ (with $k_y = 0$ from axisymmetry). 
The general features of this curve can be explained by examining which two of the three terms in Eq.\,\eqref{eq:resonances} are largest, with the $k_z w_z$ term always significant in the large $\zeta$ regime -- i.e.\ far from the midplane -- considered here.
When $k^2_x \ll k^2_z$, the resonant vertical wavenumber becomes constant, that is $k_{z} \rightarrow \varOmega_0 / w_z$.
For small $k_x$, but non-zero, the approximate solution follows as $k \sim \varOmega_0/w_z$.
When $k_x \gg k_z$ the Epicyclic frequency goes to zero and the solution becomes $k_{z} \rightarrow -k_x w_x/w_z$.  \citetalias{Squire2018} refer to this regime as ``double-resonant."
The transition between the regimes occurs near $k_x \sim 2\ \varOmega_0/w_z$.  

\citetalias{Squire2018} give simple analytic expressions for the resonant DSI growth rate, i.e. the growth rate when \Eq{eq:resonances} is satisfied, to leading order in both $\epsilon \ll 1$ and $T_{\rm s} \ll 1$.
We verified that the \citetalias{Squire2018} expressions held in the appropriate regimes. 
For example, the triangle in Fig.\,\ref{fig:fig1} shows the growth rate for a resonant angle of $\theta_k = \tan^{-1}(k_x/k_z) = 70^{\circ}$. 
We found a value of $\sigma(\theta_k)=0.023\varOmega_0$, in excellent agreement with the leading order analytic solution (Eq.\,32 in \citetalias{Squire2018}) which scales as $\sigma/\varOmega_0 \simeq \sqrt{\epsilon} k_x/k$.  

As $k_x$ increases, \citetalias{Squire2018} found that $\sigma/\varOmega_0 \simeq (\epsilon T_{\rm s}k_x \eta r_0)^{1/3}$ in the double-resonant regime.
In this regime, the DSI growth rates increase without limit towards smaller scales, which can only happen because small-scale dissipation is ignored.
In our example, the growth-rate is $\sigma = 0.45\varOmega_0$ at $k_x\eta r_0 \simeq 10^{4}$.
This gives a growth-rate 40 times larger than the settling rate, and, as mentioned, larger values can be obtained by increasing the value of $k_{x {\rm max}}$, which in absence of dissipation is an arbitrary choice. 

As suggested by \citetalias{Squire2018}, this rapid linear growth could have implications for planetesimal formation during the settling of small dust particles.
However, as we will show in Section\,\ref{sec:NonLinear}, strong clumping of small grains is absent.
Furthermore, even weak background turbulence significantly reduces even the linear growth, as shown in Section \ref{sec:Turbulent_Background}.

\subsection{Particle-size distribution}
\label{sec:poly}

We now generalize our study of the linear DSI to include a range of particle sizes.  We assume a power-law distribution of particle-size $a$, with a number density $N(a) \propto a^{s}$ fixing $s = -3.5$ as the index of differential number counts, the standard result for collisional equilibrium \citep[][]{Dohnanyi1969}. 
As previously noted, we neglect small gas density fluctuations so that particle size is proportional to $T_{\rm s}$.
Integrating the mass in this size distribution between neighboring dust species gives the dust-to-gas mass ratio of $k$th bin as
\begin{equation}
\epsilon_k = \epsilon\, \displaystyle{ \frac{T^{4+s}_{\rm s k+1} - T^{4+s}_{\rm s k}}{T^{4+s}_{\rm s, max} - T^{4+s}_{\rm s, min}}}.
\end{equation} 
Larger particles contribute more to the integrated dust mass since $4+s>0$.  
The Stokes numbers, $T_{{\rm s} k}$ are sampled on a log-uniform scale between $T_{\rm s, min} = 10^{-5}$ and $T_{\rm s, max}$, which we vary from $[10^{-4}, 0.5]$.
Each species, $k$, is given the maximum Stokes number of the corresponding bin, i.e. $T_{{\rm s} k +1}$.  This  bin edge approximation, like other aspects of the discretization, converges with the number of species.

We vary the total dust-to-gas mass ratio as $\epsilon \in [10^{-3},1]$. 
We solve the linear problem defined by Eqs.\,\eqref{Eq:linear_beg}-\eqref{Eq:linear_end} and search for the fastest growth rate over a domain $[k_{x,\min}, k_{x,\max} ] = [k_{z,\min} , k_{z,\max}] = 2\pi/H [1, 10^{3}]$, discretized in $256\times256$ mesh points.
As with the single-species case, the fastest growth was at the largest $k_x = 2 \pi 10^{3}/H$.
The choice of the maximum wavenumber is arbitrary in the absence of dissipation, however, it does affect the main results discussed below.

In Fig.\,\ref{fig:fig2} we compare the maximum growth rates to the settling rates,  as a function of $\epsilon$ and $T_{\rm s, max}$. The left panel corresponds to the case of a mono-disperse dust population, while the center panel considers $128$ dust-species.   
We discuss convergence properties in detail below but first address growth rates.

In absence of a sustained particle inflow, the DSI growth rates must at least be faster than settling rates for growth to occur before solids approach the midplane.  Moreover, growth must occur several times faster if non-linear saturation is to be achieved.  Fig.\,\ref{fig:fig2} shows that smaller solids (i.e.\ distributions with smaller $T_{\rm s, max}$) meet this growth condition more readily because they settle more slowly.  In principle, faster growth is possible for larger wavenumbers. However, as we shall show in  Section\,\ref{sec:Turbulent_Background}, the role of diffusion on small scales sharply curtails growth rates.

The right panel of Fig.\,\ref{fig:fig2} examines the convergence of growth rate with the number of dust-species, for the case of $\epsilon = 0.02$ and several $T_{\rm s, max}$ values.   
With the exception of the larger particle case with $T_{\rm s, max}=0.5$, convergence is achieved for $N \gtrsim 32$ species. 
In addition, we find that larger dust-to-gas mass ratios, $\epsilon > 0.05$, ensure convergence of the growth-rate for  $T_{\rm s, max}=0.5$ with $N\gtrsim100$ species.

The convergence of DSI growth rates (with the number of particle species) differs from the standard SI case  \citepalias{Krapp2019}.   
Compared to the DSI case above (with $\epsilon = 0.02$), the SI growth rates do not converge (up to $N = 2048$ species) for  $T_{\rm s, max} \geq 10^{-2}$ (see  top row of Fig.\, 4 form \citetalias{Krapp2019} for low $\epsilon = 0.01$ and $0.1$).  
In other words, when resolving a size distribution at low $\epsilon$ values, the DSI converges for larger $T_{\rm s, max}$ and with significantly fewer size bins, that is $N \simeq 16 - 32$.  
We stress that the SI is most significant for particle clumping at larger $\epsilon \geq 0.5$ \citep{Johansen&Youdin2007}, where \citetalias{Krapp2019} found good convergence for the growth-rate for $T_{\rm s,max} \leq 0.1$, but not for $T_{\rm s, max}=1$ (similar to the $T_{\rm s, max}=0.5$ DSI case studied here).  
 
Convergence seems to be achieved when the distance between the resonances is shorter than their width.
When increasing the number of species, the width of the existing resonances is divided into new resonances.
This width scales with the dust-to-gas mass ratio of the species, $\epsilon_j$, and thus it decreases as more species are included.  DSI resonant modes have a much broader width in comparison with the SI at low $k_x$ values, which helps the resonances to overlap once a few species are included ($N \simeq 16$).
Moreover, for the DSI at large $k_x$ (where the double-resonant condition ${\bf k} \cdot {\bf w} = 0$ is satisfied), the resonant wavenumbers of each species are independent of $T_{{\rm s}j}$.
This is because both the radial and vertical drift speeds scale linearly with the Stokes number (to leading order).
Besides, the resonance width is less sensitive to $\epsilon_j$ at these scales.
Thus, at a fixed wavenumber, all the particle species are still very close to resonance.
In contrast, for the SI the resonances are more widely separated, and thus the distribution of particle sizes will span both resonant and non-resonant parts of parameter space.
Consistent with this picture, choosing a larger $\epsilon$ value increases the resonant width in the SI, eventually allowing overlap and thus convergence for $T_{\rm s, max} \lesssim 0.1$. 

We conclude that the inclusion of a standard particle-size distribution has a minor impact on the linear growth of the DSI.
Furthermore, the time and length scales of the instability are comparable to those from the monodisperse case.
We will then focus on the non-linear dynamics of the DSI including one dust-species in our simulations and leave the multi-species case for future work.

\begin{figure*}
	\includegraphics[]{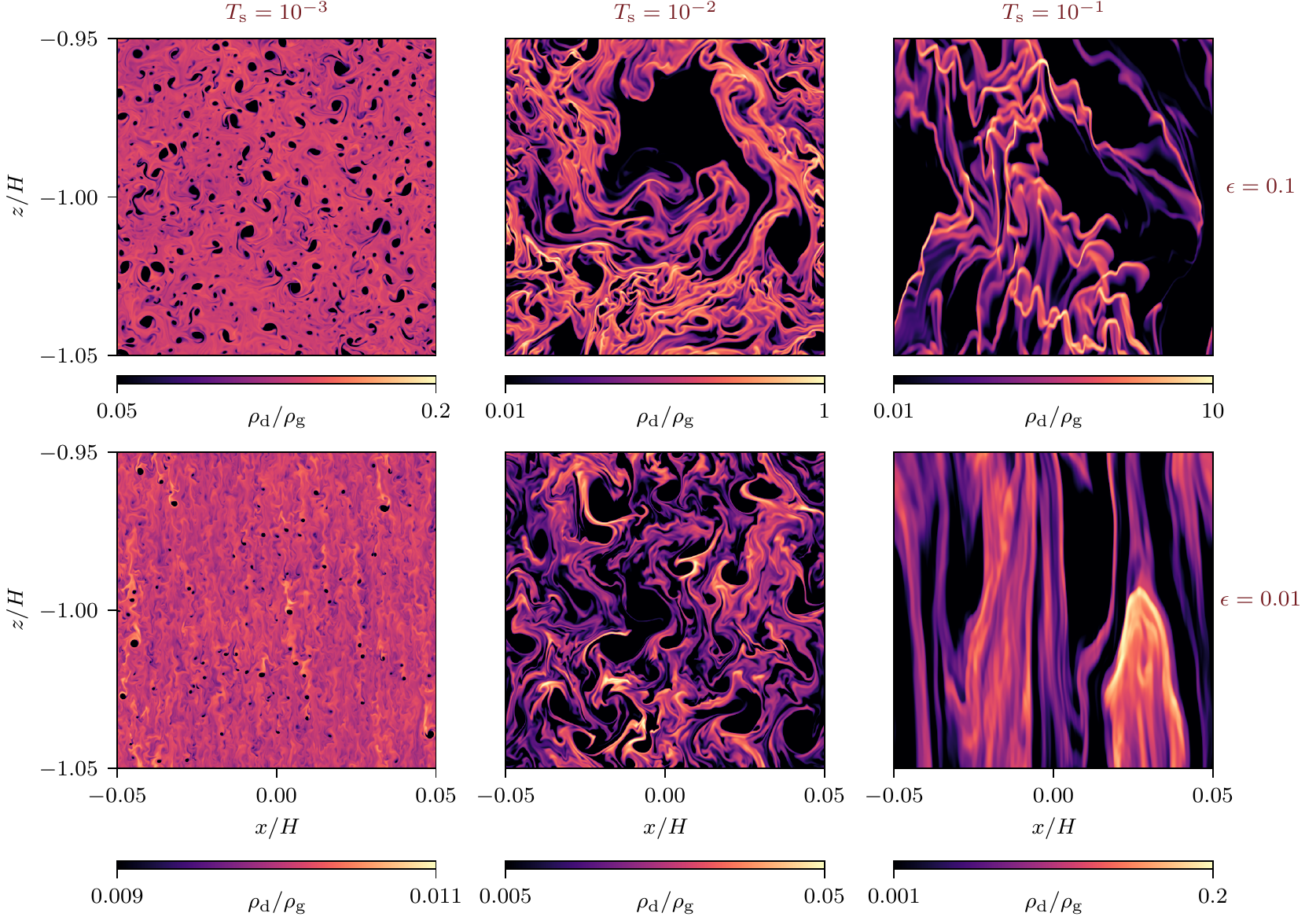} \caption{Dust-density snapshots during the saturated state of local, non-linear DSI simulations of dust settling at $H$ away from the midplane. Top and bottom rows correspond to dust-to-gas mass ratios of $\epsilon=10^{-1}$ and $\epsilon=10^{-2}$, respectively. 
		From left to right the Stokes number are $T_{\rm s} = 10^{-3}$, $T_{\rm s} = 10^{-2}$ and $T_{\rm s} = 10^{-1}$. 
		The colorbars for the dust density (scaled to the average gas density) vary for each panel to capture the range of density fluctuations.  The maximum density (and other details) are in Table\,\ref{tab:runs} for these runs (the ``s2D" series) and others.
		Only the top right case ($T_{\rm s}=10^{-1}$ and $\epsilon=10^{-1}$) produces significant clumping ($\rho_{\rm d} \gtrsim 10 \rho_{\rm g}$).  However the clumping time is longer than the settling time for this case (see \Fig{fig:fig4}), so the clumping seen in our local simulations would not occur in a real disc (with vertical structure).
		The prediction that DSI would produce strong particle clumping is not supported by these simulations, with very weak clumping for the most interesting case of Solar abundances ($\epsilon \sim 0.01$ at $H$) and small solids.	}	\label{fig:fig3}
\end{figure*}

\begin{table*}
	\centering
	\caption{Numerical Simulations with FARGO3D}
	\label{tab:runs}
	\setlength{\tabcolsep}{5.5pt}{\begin{tabular}{lllccccccl}
			\hline\hline
			Run  & $(L_x \times L_z)/H^2 $  & $N_x \times N_z \times N_y$ & $\epsilon$ & $T_{\rm s}$ & $\alpha_{xy}$  & $\delta v^2_{{\rm g}y}/c^2_{\rm s}$ & $\delta v^2_{{\rm g}x}/c^2_{\rm s}$ & $\delta v^2_{{\rm g}z}/c^2_{\rm s}$ & ${\rm max}(\rho_{\rm d})/\rho^0_{\rm g}$\\
			\hline
			s2De1T1 & $ 0.1  \times 0.5$   & $1024 \times 2560 \times 1$   & $10^{-1}$  & $10^{-1}$  & $ 4.4 \times 10^{-4}$   & $5.9  \times 10^{-4}$   & $4.9  \times 10^{-3}$  & $5.7  \times 10^{-3}$ & $22.44$\\
			s2De1T2 & $ 0.1  \times 0.1$   & $1024 \times 1024 \times 1$   & $10^{-1}$  & $10^{-2}$  & $ 2.6 \times 10^{-4}$   & $5.0  \times 10^{-4}$   & $3.0  \times 10^{-3}$  & $3.5  \times 10^{-3}$ & $1.302$\\
			s2De1T3 & $ 0.1  \times 0.1$   & $1024 \times 1024 \times 1$   & $10^{-1}$  & $10^{-3}$  & $ 5.7 \times 10^{-6}$   & $5.0  \times 10^{-5}$   & $1.1  \times 10^{-4}$  & $1.5  \times 10^{-4}$ & $0.14$ \\
			s2De2T1 & $ 0.1  \times 0.5$   & $1024 \times 2560 \times 1$   & $10^{-2}$  & $10^{-1}$  & $ 1.2 \times 10^{-5}$   & $8.0  \times 10^{-5}$   & $1.9  \times 10^{-4}$  & $3.1  \times 10^{-4}$ & $0.225$\\
			s2De2T2 & $ 0.1  \times 0.1$   & $1024 \times 1024 \times 1$   & $10^{-2}$  & $10^{-2}$  & $ 1.1 \times 10^{-5}$   & $7.0  \times 10^{-5}$   & $2.2  \times 10^{-4}$  & $2.6  \times 10^{-4}$ & $0.062$\\
			s2De2T3 & $ 0.1 \times 0.1$   & $1024  \times 1024 \times 1$   & $10^{-2}$  & $10^{-3}$  & $ 5.0 \times 10^{-8}$   & $3.0  \times 10^{-5}$   & $3.0  \times 10^{-5}$  & $3.0  \times 10^{-5}$ & $0.011$\\
			n2De1T2 & $ 0.1  \times 0.1$   & $2048 \times 2048 \times 1$   & $10^{-1}$  & $10^{-2}$  & $ 1.9 \times 10^{-4}$   & $3.4  \times 10^{-4}$   & $2.2  \times 10^{-3}$  & $2.6  \times 10^{-3}$ & $3.172$\\
			n2De1T3 & $ 0.01  \times 0.01$   & $1024 \times 1024 \times 1$   & $10^{-1}$  & $10^{-3}$  & $1.1 \times 10^{-5}$   & $4.1 \times 10^{-5}$   & $3.0\times 10^{-4}$  & $2.5 \times 10^{-4}$ & $0.263$\\
			\hline
			c2D    & $ 0.1 \times 0.1$   & $256 \times 256 \times 1$    & $10^{-1}$  & $10^{-2}$  & $1.9 \times 10^{-4}$     & $4.4 \times 10^{-4}$    & $2.0 \times 10^{-3}$  & $1.9 \times 10^{-3}$  & $0.368$\\ 
			c3D   & $ 0.1 \times 0.1 \times 0.1H$ & $256 \times 256 \times 256$ & $10^{-1}$  & $10^{-2}$  & $6.2 \times 10^{-6}$     & $6.3 \times 10^{-5}$    & $4.5 \times 10^{-5}$  & $1.0 \times 10^{-4}$  & $0.23$\\
			\hline
		\end{tabular}
	}
	{\hspace{5pt} $\alpha_{xy}$, $\delta v^2_{{\rm g}x}/c^2_{\rm s}$, $\delta v^2_{{\rm g}y}/c^2_{\rm s}$,  $\delta v^2_{{\rm g}z}/c^2_{\rm s}$, and ${\rm max}(\rho_{\rm d})/\rho^0_{\rm g}$ correspond to the time averaged values obtained between times $60\varOmega^{-1}$ and $80\varOmega^{-1}_0$, except for the runs c2D and c3D where the time average was obtained between times $50\varOmega^{-1}$ and $60\varOmega^{-1}_0$, and run n2De1T3 where time average was obtained between times $10\varOmega^{-1}$ and $30\varOmega^{-1}_0$. For the 3D run the domain size is expanded in the azimuthal direction with a size of $L_y=0.1H$, while for the 2D runs the value of $L_y$ is omitted. }
\end{table*}

\section{Numerical Simulations}
\label{sec:NonLinear}

We study the nonlinear evolution of the DSI in both 2D and 3D shearing boxes using the multi-fluid version of the publicly available code FARGO3D\footnote{fargo3d.bitbucket.io, fargo.in2p3.fr} \citep{Benitez-Llambay2016,Benitez-Llambay2019} using the FARGO orbital advection scheme \citep{Masset2000}.
These simulations allow us to measure the strength of dust concentration in the saturated state and the level of turbulence generated by the DSI.
We focus on dust-to-gas mass ratios and Stokes numbers where the saturation time is faster than the settling time, and relatively strong clumping develops. 

We carry out single dust-species runs, where density, velocity and Stokes number are denoted by $\rho_{\rm d}$, ${\bf v}_{\rm d}$ and $T_{\rm s}$, respectively.\footnote{
We omit the dust species subscript $j=1$ used in the previous section.}
We adopt code units such that $\varOmega_0 =1$, $r_0 = 1$, and the gas initial density is $\rho^0_{\rm g} = 1$. 
The initial dust density is defined as $\rho^0_{\rm d} = \rho^0_{\rm g}\epsilon$, and the radial and azimuthal velocities are initially set to the steady-state solutions shown in Eqs.\,\eqref{Eq:steady_state_gas}\,-\,\eqref{Eq:steady_state_dust}, with the additional shear velocity in the azimuthal direction. 
As described in Appendix\,\ref{sec:linearized}, we adopt a coordinate system where $v_{{\rm g}z}=0$ and $v^0_{{\rm d}z} = T_{\rm s}z_0\varOmega_0$. 
We set the final integration time to $80\varOmega^{-1}_0$, which is long enough to capture saturation for the slowest settling dust species tested.

The domain sizes, number of cells, $N$, and values for $\epsilon$ and $T_{\rm s}$ are given in Table\,\ref{tab:runs} for reference. 
Because of our choice of units, and the fact that we fix $\eta r_0 = (H/r_0)^2$ and $z_0=H$ (see Section \ref{sec:Equations}), the sound speed and external accelerations are defined through the parameter $h_0 = H/r_0 = 10^{-3/2}$ \citep[see e.g.,][]{Benitez-Llambay2019}.
To excite the instability, we add white noise to the velocities with an amplitude of $10^{-2}c_{\rm s}$. 
With these relatively large initial perturbations, the instability rapidly transitions to non-linear growth. Smaller initial noise perturbations (considered in Section\,\ref{sec:spectra})) produce an extended linear-growth phase but a similar non-linear saturated state.  The saturation timescales provided in Section\,\ref{sec:2D} apply to the larger (default) level of initial noise, and thus could be  underestimates.
To properly recover the linear growth we require a minimum of $16$ cells per wavelength of the fastest mode\footnote{The dispersion relation of the DSI implies a faster growth with higher resolution. However, as numerical diffusion prevents the growth at the grid resolution, we assume the fastest mode with $k_x \simeq 2\pi/ (16 \Delta x)$, where $\Delta x$ the radial distance between two consecutive grid points based on the results obtained in Appendix\,\ref{sec:Test}.}.

\subsection{2D Simulations}
\label{sec:2D}
In Fig.\,\ref{fig:fig3} we show the dust densities at time $60 \Omega^{-1}_0$ for the six runs of the series s2D (see Table\,\ref{tab:runs}). 
Upper and lower panels correspond to different dust-to-gas mass ratios, with the lower $\epsilon = 0.01$ corresponding to standard Solar abundances (still perhaps an overestimate if size distributions are considered) and the higher $\epsilon = 0.1$, which is chosen mainly to attempt to trigger stronger concentrations, but would require an unspecified mechanism to concentrate dust not just in the midplane, but up to $z_0 = H$.  
Columns from left to right show the results for different Stokes number. 
For all of the cases tested, we find only moderate enhancements in the dust-gas ratio in the saturated state.
Starting at $\epsilon = 0.01$, this ratio remains near $\sim 1\%$ for $T_{\rm s} = 10^{-3}$ increasing to $\sim 20 \%$ at $T_{\rm s} = 10^{-1}$, with larger initial $\epsilon$ values producing stronger clumping for all Stokes numbers.  
In agreement with the assumption of low compressibility of Section\,\ref{sec:Equations}, time average of $({\rm max}(\rho_{\rm g}) - \rho^0_{\rm g})$ are below one percent for $\epsilon=0.01$ and increase up to eight percent for $\epsilon=0.1$.

Properties of the $T_{\rm s}=10^{-1}$ simulations lead us to conclude that the strongest clumping seen in this case is not achievable in practice.  Thus (of the cases considered) $T_{\rm s}=10^{-2}$ would give the strongest clumping in practice.  To explain, first note that the  $T_{\rm s}=10^{-1}$ simulations used a more extended vertical domain of $\Delta z \simeq 0.5H$, compared to  $\Delta z \simeq 0.1H$ for the other runs.  The motivation was to fit both the rapidly growing linear modes with large $\lambda_z$ (see Fig.\,\ref{fig:add1}, described below) and also more of the vertically elongated nonlinear structures seen in Fig.\,\ref{fig:fig3}.
In such a tall box, the use of a constant vertical acceleration should be replaced by the stratified potential and, correspondingly, a non-periodic vertical boundary condition.
While beyond the scope of this work, the buoyancy in a stratified model should inhibit the clumping seen in our $T_{\rm s}=10^{-1}$ local simulations.  
Second, and more definitively, the clumping seen for $T_{\rm s}=10^{-1}$ is too slow relative to particle settling, as we describe next.
Thus we caution that the clumping in our $T_{\rm s}=10^{-1}$ simulations should not be taken at face value.

\begin{figure*}
	\includegraphics[]{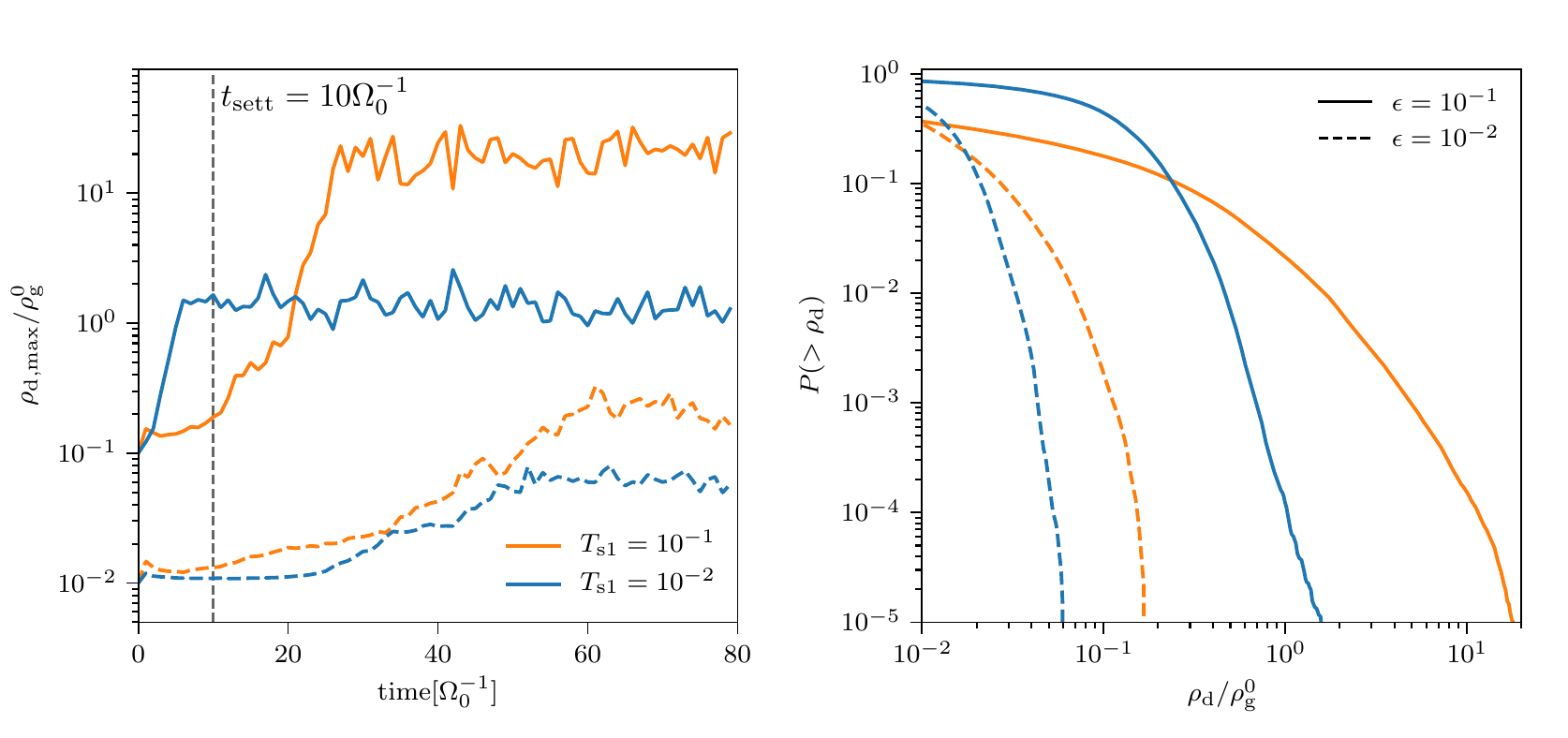} \caption{Analysis of particle clumping for the s2D runs of \Fig{fig:fig3} and Table \ref{tab:runs} (omitting  $T_{\rm s}=10^{-3}$ runs due to negligible clumping).
	Blue and orange lines correspond to $T_{\rm s}=10^{-2}$ and $T_{\rm s}=10^{-1}$ (respectively), while dashed and solid lines correspond to $\epsilon = 10^{-2}$ and $\epsilon = 10^{-1}$, respectively. 
	\emph{(Left)} Time evolution of the maximum dust-density.  The vertical dashed line shows the settling time for $T_{\rm s} = 0.1$, showing that clumping in these runs takes longer than a settling time.  The settling time for $T_{\rm s} = 0.01$ is $100/\varOmega_0$, and thus (weak) clumping occurs faster than settling in this case.
	\emph{(Right)} The cumulative probability distribution of particle density.  }	\label{fig:fig4}
\end{figure*}

\begin{figure}
	\includegraphics[width=\columnwidth]{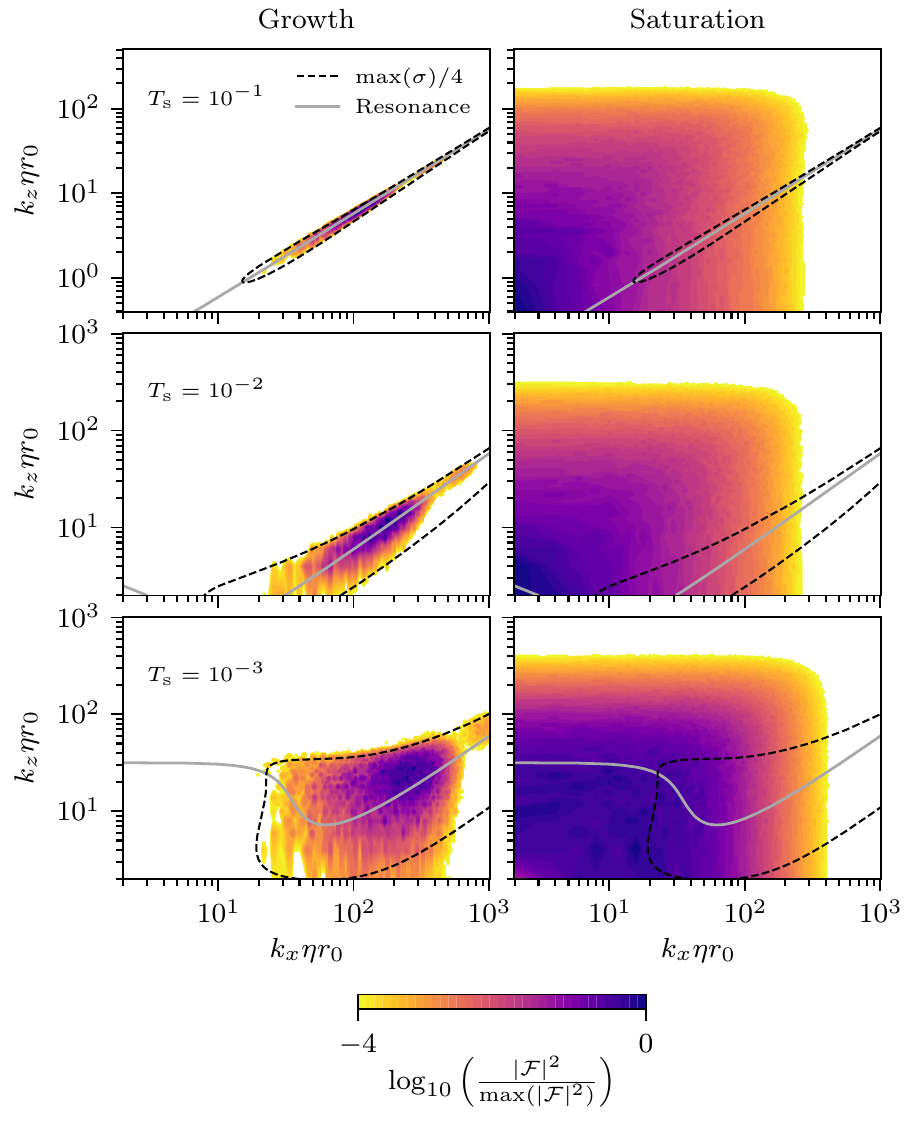} 
	\caption{Spectra of the dust-density for different Stokes numbers during the growth and saturation phases. From top to bottom the runs are s2De1T1, s2De1T2 and s2De1T3. $\mathcal{F}$ is the 2D Fourier transform of $\rho_{\rm d}/\rho^0_{\rm g}$. The colorscale is truncated at $|\mathcal{F}|^2/{\rm max}(|\mathcal{F}|^2) \leq 10^{-4}$. The solid grey line corresponds to the resonant modes where the power should be concentrated during the linear growth phase. The dashed black contours correspond to an approximate resonant width, where the growth rates are within one quarter of the maximum growth rate. 
	During the saturated phase the maximum of the power is not concentrated along the resonance.}
	\label{fig:add1}
\end{figure}

Fig.\,\ref{fig:fig4} shows the time evolution of the maximum dust-density, which provides a useful estimate of the saturation timescale, $t_{\rm sat}$.
For $T_{\rm s}=10^{-1}$ this timescale corresponds to $t_{\rm sat} \sim 20 \varOmega^{-1}_0$, twice the settling time. 
Thus $T_{\rm s}=10^{-1}$ particles would settle to the mid-plane before the instability saturates, and the relatively strong clumping in local models is not consistent with these vertically global considerations. 
By contrast, in simulations with $T_{\rm s}=10^{-2}$ clumping saturates much faster than the settling time, for both $\epsilon$ values.
When $\epsilon = 10^{-1}$, saturation occurs within one vertical dust-crossing time,
$t_{\rm cross} = L_z/v^0_{{\rm d}z}$.
For a more realistic solid fraction for young PPDs, $\epsilon = 10^{-2}$, saturation is still reached within one settling time, but this now corresponds to a few vertical crossing times.
Therefore the fully saturated turbulent regime of the DSI may be only obtained if there is a substantial inflow of particles at $z_0 \geq H$. 
To properly capture the DSI in stratified numerical simulations requires resolution comparable to our local simulations as well as a vertical domain that spans above and below $H$, where DSI growth is optimized (assuming that sufficient amount of dust is stirred to $H$).
\begin{figure}
	\centering
	\includegraphics[width=\columnwidth]{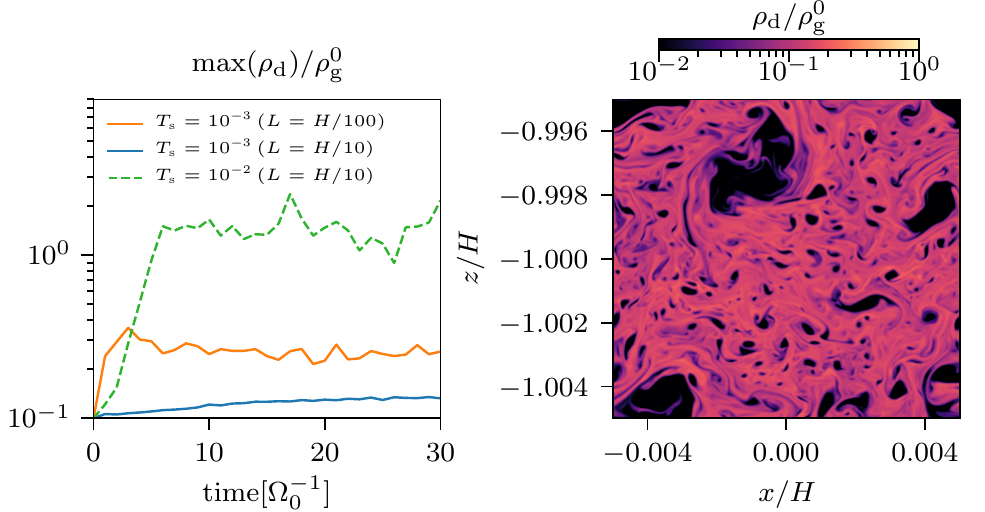}
	\caption{{Comparison between simulations with different box sizes for the case with $T_{\rm s}=10^{-3}$ and $\epsilon=0.1$. The left panel shows the maximum dust density as a function of time for the runs s2De1T3 (solid blue line), n2De1T3 (solid orange line), and for comparison the $T_{\rm s} = 10^{-2}$ case from Fig.\,\ref{fig:fig4}} (run s2De1T2, dashed green line). The right panel illustrates the lack of strong clumping for $T_{\rm s} = 10^{-3}$ in a snapshot of the dust density at time $30\varOmega^{-1}_0$ for run n2De1T3.  
	Decreasing the box size from $L=H/10$ to $L=H/100$ results in a maximum density increase of roughly a factor $2$ (see Table\,\ref{tab:runs} for reference). }
	\label{fig:compare}
\end{figure}
To better compare with previous work \citep{Johansen2007, Bai2010, Benitez-Llambay2019}, we also characterize the DSI clumping properties via the cumulative-particle-density distribution (CPD), shown in the right panel of Fig.\,\ref{fig:fig4}. 
We compute the CPD by first defining 400 log-spaced density bins in a range of $[10^{-1}\epsilon, 2\times10^{2}\epsilon]$ and then counting the cells whose density value is above the specified threshold. 
All distributions are normalized such that they integrate to unity. We omit runs with $T_{\rm s}=10^{-3}$ as they show maximum dust-densities less than double that of the initial conditions. 
We compare the CPD for Stokes numbers $T_{\rm s}=10^{-2}$ and $T_{\rm s}=10^{-1}$ for both  values of $\epsilon$.
The CPDs clearly show $\epsilon$ dependence that becomes more prominent for larger particles. This finding is consistent with the observed large scale filaments with denser clumps as the dust-to-gas mass ratio increases from $\epsilon=0.01$ to $\epsilon=0.1$.
\begin{figure*}
	\includegraphics[]{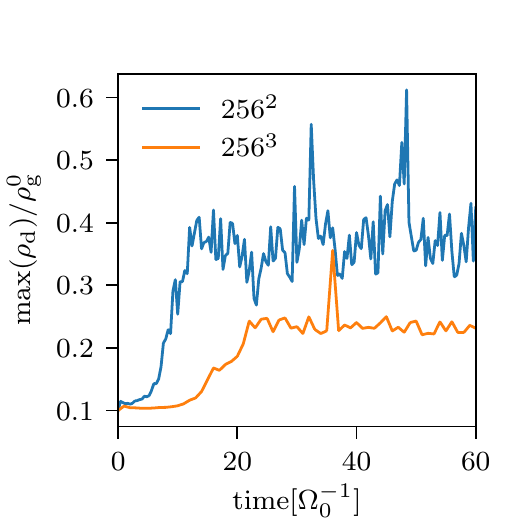} 
	\includegraphics[]{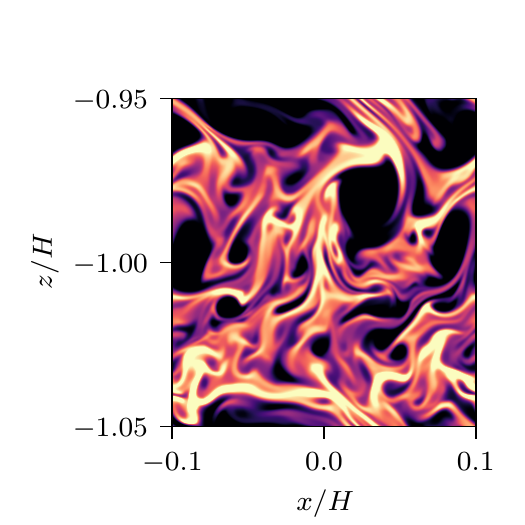} \hfill
	\includegraphics[height=0.2\textheight]{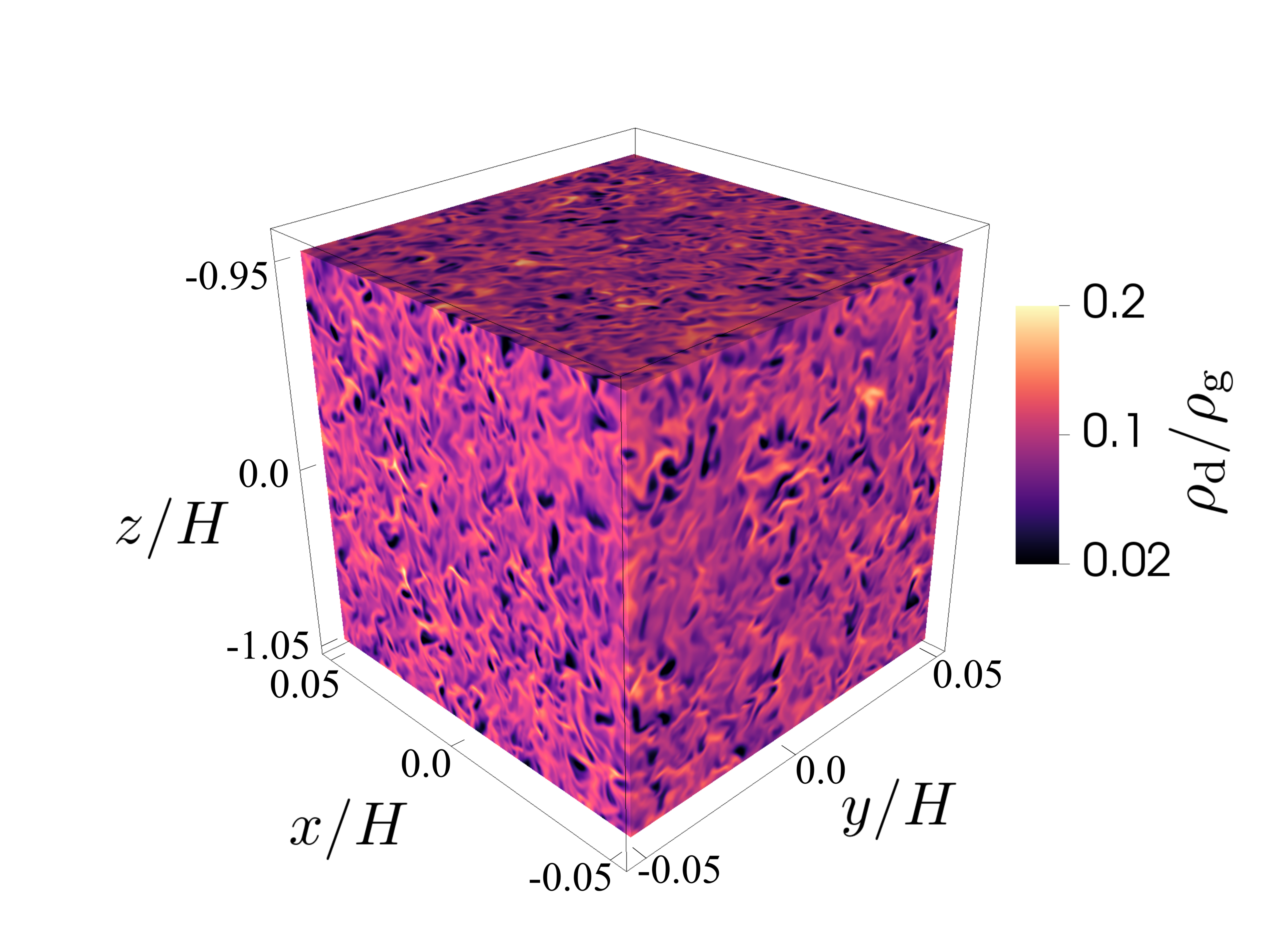} 
	\caption{A comparison of particle clumping in 2D vs. 3D simulations at the same resolution (runs c2D and c3D in Table \ref{tab:runs}).  \emph{(Left)}  Time evolution of the maximum dust density, showing that clumping in 3D is both weaker on average and more constant in time than in 2D. \emph{(Center and Right)} Snapshots of the 2D and 3D simulations, respectively, after $50\varOmega^{-1}_0$, using the same colorbar.  In 3D, turbulent scales are significantly smaller than in 2D (despite identical resolution) and clumping is a factor of $\sim 2$ weaker.}
	\label{fig:fig6}
\end{figure*}

\subsubsection{Turbulence in the saturated state}
\label{sec:saturation}
In addition to measuring clumping properties, we can also characterize the self-generated turbulence due to the DSI in the saturated state. 
We measure the dimensionless Reynolds stress component, $R_{xy}$, together with the gas velocity fluctuations. 
The dimensionless Reynolds stress is calculated as follows
\begin{align}
R_{xy} &=  \frac{ \int \rho_{\rm g}v_{{\rm g}x} (v_{{\rm g}y}-qx\varOmega_0)  {\rm d}V} { c^2_{\rm s} \int \rho_{\rm g} {\rm d}V },
\end{align}
whereas the velocity fluctuations are obtained after subtracting the mean velocity, $\bar{\bf v}_{\rm g}$, that is, $\delta {\bf v}_{\rm g} = {\bf v}_{{\rm g}} - \bar{\bf v}_{{\rm g}}$. 
The dimensionless Reynolds stress is usually adopted as a proxy for the turbulent viscosity as $\alpha_{xy} \equiv \langle R_{xy}\rangle_{T}$, where the brackets denote time average.
The gas velocity fluctuations can be used to estimate the particle diffusion coefficient, $D_{\rm DSI}$, \citep{YoudinLithwik2007}. 
In this case, we assume that the inverse of the turnover time of the largest eddy is of the order of the dynamical frequency, thus  $D_{\rm DSI} \sim \delta v_{\rm g}^2\varOmega_0$.
Note that in our pressureless fluid approach adopted for the dust-species, a direct estimation of dust diffusion by tracing the particle orbits is prohibited.

In Table\,\ref{tab:runs} we show the time-averaged values of $\alpha_{xy}$, $\delta v^2_{{\rm g}x}$, $\delta v^2_{{\rm g}y}$ and $\delta v^2_{{\rm g}z}$ between $60\varOmega^{-1}_0$ and $80\varOmega^{-1}_0$. 
Values for the dust velocity fluctuations are omitted because they differ from those of the gas by an order unity factor.
The saturated regime of the DSI can induce a low-to-moderate turbulent viscosity for Stokes numbers $T_{\rm s }\gtrsim 10^{-2}$, that is $\alpha_{xy} \simeq 10^{-5} - 10^{-4}$, depending on the dust-to-gas mass ratios. 
Comparable values (within a factor order unity) are obtained for $\alpha_{zy} \equiv \langle R_{zy}\rangle_{T}$, where $R_{zy}$ is the vertical-azimuthal component of the Reynolds stress. 
This is in agreement with the also similar values obtained for the radial and vertical gas velocity fluctuations.

Considering the average vertical velocity fluctuations,  we estimate turbulent diffusion coefficients $D_{\rm DSI} \sim 10^{-4} c^2_{\rm s} \varOmega_0$ and $D_{\rm DSI} \sim 10^{-5} c^2_{\rm s} \varOmega_0$, for Stokes numbers $T_{\rm s}=10^{-2}$ and $T_{\rm s}=10^{-3}$, respectively.
These values are not large enough to prevent the settling of particles because the settling timescale is faster than the diffusion timescale (across the relevant length $H$). 
Comparable values are obtained for the radial diffusion coefficient, indicating that the turbulence triggered by the DSI is isotropic on these scales.

As a consequence of this low -- but non-negligible -- particle diffusion, the DSI may have implications for the coagulation and sticking of $\rm mm$ to $\mu \rm m$-sized particles in PPDs, depending on the particle relative velocity induced by the turbulence \citep[e.g.,][]{Ormel2007, Birnstiel2016}. As we discuss Section\,\ref{sec:Turbulent_Background}, this level of turbulence also necessitates a reconsideration of the linear phase of the instability in the non-laminar case.

\subsubsection{Dust Density Power Spectrum}
\label{sec:spectra}

In Fig.\,\ref{fig:add1} we show the power spectra of the dust-density, i.e. the squared FFT amplitudes, for the simulations with $\epsilon=0.1$.  Though such high dust abundances are unlikely to occur at $z_0=H$, they produce larger density fluctuations.
To better capture the linear growth phase, we reduce the initial white noise perturbations to $\sim 10^{-6}c_{\rm s}$, much smaller than the fiducial runs.  
The non-linear outcome remains the same, but the smaller perturbations produce cleaner power spectra in the linear phase.

The left panels of Fig.\,\ref{fig:add1} show the time-averaged power during the initial growth phase of the DSI.
For $T_{\rm s} = 10^{-1}$, $T_{\rm s} = 10^{-2}$ and $T_{\rm s} = 10^{-3}$  times analyzed  are $9\varOmega^{-1}_0 - 11\varOmega^{-1}_0$, $14\varOmega^{-1}_0 - 16\varOmega^{-1}_0$ and $16\varOmega^{-1}_0 - 18\varOmega^{-1}_0$, respectively.
For all Stokes numbers, the maximum power is concentrated along the resonant condition during the linear phase.  The ``resonant width" is indicated by dashed contours where $\sigma = 0.25{\rm max}(\sigma)$, with ${\rm max}(\sigma)$ the fastest growth rate in the Fourier domain.  Power is  $|\mathcal{F}|^2 \lesssim 10^{-4} {\rm max}(|\mathcal{F}|^2)$  outside this resonant width, with ${\rm max}(|\mathcal{F}|^2)$ the maximum value.

The right panels of Fig.\,\ref{fig:add1} show the time-averaged power spectra during the saturated phase. For all Stokes numbers this average covers times $75\varOmega^{-1}_0 - 80\varOmega^{-1}_0$.  Remarkably, the power during the saturated phase no longer traces the resonant condition of the DSI.  This finding complicates efforts to develop a non-linear theory of the DSI and related RDIs. 

For $T_{\rm s}=10^{-1}$ and $T_{\rm s}=10^{-2}$, the saturated power is largest at the smallest wavenumbers, indicative of an inverse cascade.  This trend reflects the large, box-scale features seen in the corresponding snapshots in the top row of Fig.\,\ref{fig:fig3}.
By contrast, for $T_{\rm s}=10^{-3}$, the maximum power occurs on intermediate scales of $k \eta r_0 \simeq 10$.  The corresponding snapshot in Fig.\,\ref{fig:fig3} consistently shows that the dominant eddies are much smaller than the domain.  
These large-scale eddies are well resolved with $\sim 100$ cells (see Fig.
\,\ref{fig:compare} for a higher resolution simulation).

The 2D spectra also clearly indicate the damping of power on small scales.  In the linear regime, modes with  $k_x \eta r_0 \gtrsim 400$ show reduced power despite their large growth rates, consistent with numerical dissipation.  Since these scales have $< 16$ cells per (radial) wavelength the damping agrees with the linear  convergence study in Appendix\,\ref{sec:Test}.  
The saturated state reveals a combination of numerical and physical damping at small scales.  While turbulent gas motions are damped by numerical viscosity in these inviscid simulations, the turbulent motions also diffuse small scale dust concentrations.

\subsection{Comparison with 3D simulations}

We extend our analysis to 3D, including the full azimuthal dynamics for the most promising clumping case with $T_{\rm s}=10^{-2}$ and $\epsilon = 10^{-1}$ (see runs s2De1T2 and c3D in Table\,\ref{tab:runs}). 
For this 3D run, we decrease the resolution and consider a box with $256^3$ cells. Thus for comparative purposes we also add a 2D run with $256^2$.
In Fig.\,\ref{fig:fig6} we show snapshots of the dust-density for the c2D and c3D run at the integration time $50\varOmega^{-1}_0$. 
See the last two rows of Table\,\ref{tab:runs} for more on these runs.
The left panel shows the evolution of the maximum dust density for both runs, clearly, the 3D run shows a maximum dust-density reduced roughly a factor $\sim 2$.
A similar decrease was also seen by \cite{Johansen&Youdin2007} when studying an unstratified SI simulation with $\epsilon=1$ and $T_{\rm s}=10^{-1}$ (AB run). 
Turbulence properties are also affected by the inclusion of the extended azimuthal domain.
The turbulent kinetic energy is reduced in the 3D run,  by nearly one order of magnitude for the vertical component, and nearly two orders of magnitude for the radial and azimuthal components, as shown in Table\,\ref{tab:runs}). 

\subsection{Numerical Convergence}
\label{sec:conv}

We carry out a brief study of the non-linear convergence of 2D simulations. First we vary the number of grid cells at fixed box size for $T_{\rm s}=10^{-2}$ and $\epsilon=10^{-1}$ (runs s2De1T2, n2De1T2  and c2D), a case with fast saturation times and moderate clumping.   
We also examine higher resolution in a smaller domain for $T_{\rm s}=10^{-3}$ and $\epsilon=0.1$, shrinking the box lengths in  run n2De1T3 by $1/10$th of s2De1T3.  For both experiments, the turbulence properties of the gas and dust converge, while the maximum dust concentration increases with resolution (see Table \ref{tab:runs}).  Interestingly, the resolution dependence of particle clumping varies with Stokes number. For $T_{\rm s}=10^{-2}$  the maximum dust density scales linearly with the resolution, while for $T_{\rm s}=10^{-3}$, increasing resolution by a factor of 10 only increases clumping by a factor $\sim 2$.

Fig.\,\ref{fig:compare} shows the maximum dust-density as a function of time for the box-size study, together with a snapshot of the dust-density for the small box ($L_x =L_z = H/100$) and therefore high resolution case (run n2De1T3.) 
The comparison run with $T_{\rm s} = 10^{-2}$ is in the larger box ($L_x=L_z=H/10$), and all three simulations  have the same number of grid cells ($1024^2$).  
According to linear theory, a similar evolution of the instability should be recovered with the small box run with $T_{\rm s} = 10^{-3}$ and the larger box run with $T_{\rm s} = 10^{-2}$. 
This is interpreted as a consequence of the growth rate scaling with the product $k_x \eta r_0 T_{\rm s}$.
However, the $T_{\rm s} = 10^{-2}$ run gives stronger clumping, confirming that non-linear dynamics is agnostic to the scaling obtained in the linear regime, and moreover,  the maximum dust-density values seen for larger $T_{\rm s}$ is not a resolution-dependent artifact. 
In other hand, the stronger clumping in the small vs.\ large boxes for $T_{\rm s} = 10^{-3}$ (orange and blue solid lines in Fig.\,\ref{fig:compare}) is consistent with the resolution dependence already noted.

The resolution dependence of particle clumping was also seen in the FARGO3D simulations of the (unstratified) SI \citep{Benitez-Llambay2019} for the particular run AB ( $T_{\rm s}=0.1$ and $\epsilon=1$), which also showed  particle clumping that increased with resolution.  
By contrast, ``hybrid" simulations (with gas as a fluid and Lagrangian particles) using ATHENA show good convergence of clumping with resolution for the same run \citep{Bai2010}.   FARGO3D gave the best agreement with the converged ATHENA result for an intermediate resolution of $1024^2$, emphasizing that higher resolution with two fluid methods is not necessarily more accurate. Hybrid methods are expected to show improved convergence at small scales, due to the inclusion of a particle velocity dispersion (i.e.\ crossing trajectories).

We believe our use of the two-fluid method is well justified scientifically, i.e. beyond the practical issue of lower computational cost.
First,  the \cite{Benitez-Llambay2019} results for SI indicate that FARGO3D simulations can not only reproduce linear growth very accurately, but also determine whether strong clumping occurs; this validation makes it a good choice for an initial study of the non-linear behavior of the DSI.  Second,  enhanced clumping at very small scales -- whether physical or numerical -- would be prevented by the presence of small scale turbulent diffusion, as we show (for linear growth) in Section\,\ref{sec:Turbulent_Background}. 

We are thus reasonably confident that the weak clumping inferred from our simulations is a real issue for the DSI and not a numerical artifact.  This confidence grows with even weak disc turbulence.   Nevertheless, our results should be tested against different, and ideally hybrid, methods.

\begin{figure*}
	\includegraphics[]{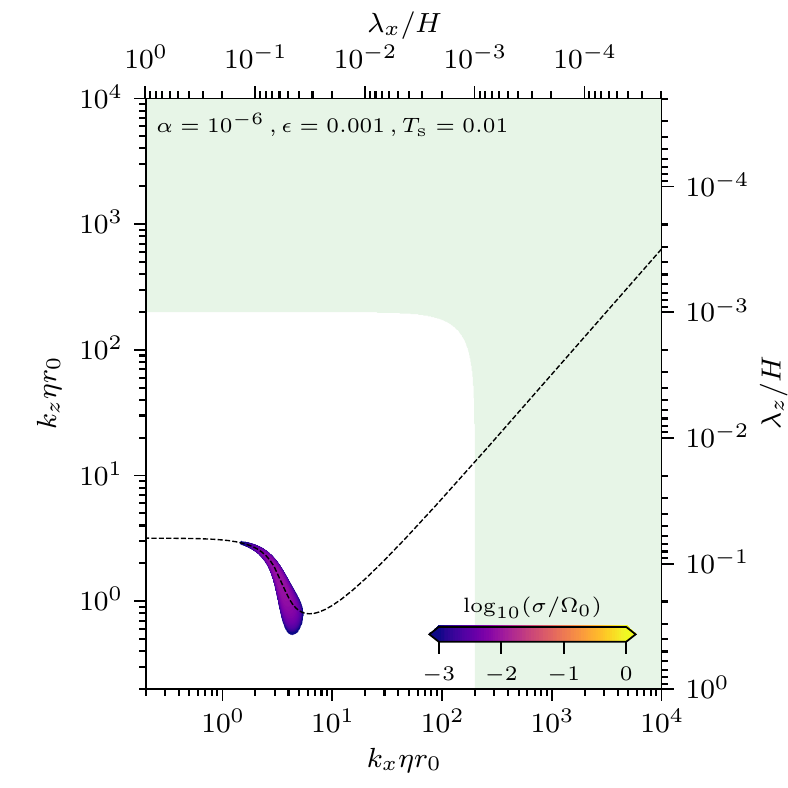}
	\includegraphics[]{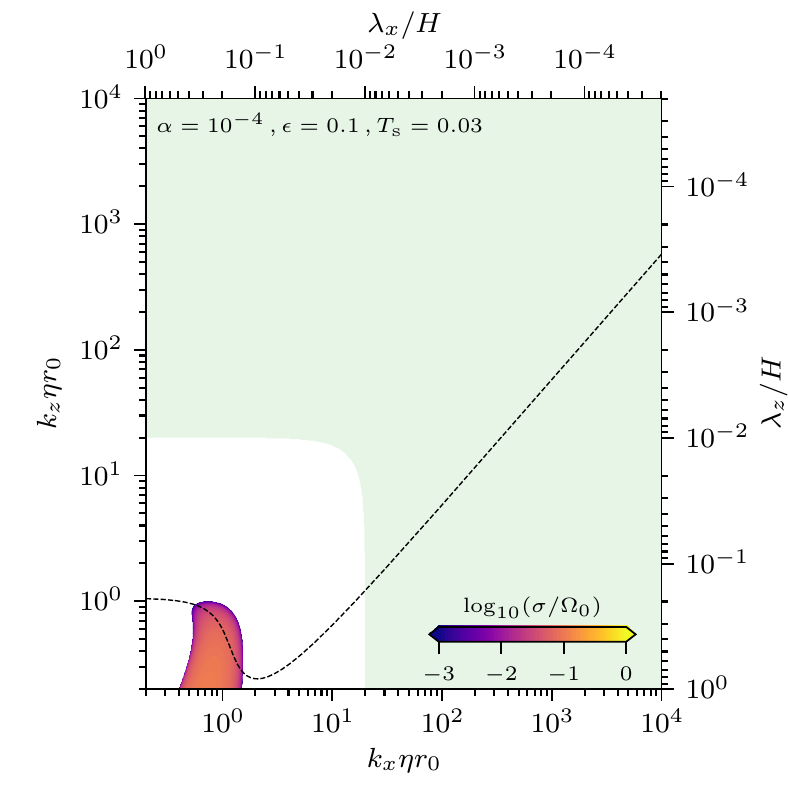}
	\caption{ Growth rates for the settling instability with background turbulence. \emph{(Left)}  Weak turbulence ($\alpha = 10^{-6}$) is added to the laminar case considered in Fig.\,\ref{fig:fig1}, showing that turbulence reduces the growth rates and the range of unstable wavenumbers.  \emph{(Right)} A different set of parameters that is near the stability boundary (which corresponds to the orange dashed line in \Fig{fig:fig8}).
	The green shaded region indicates scales smaller than the turbulent outer scale, i.e. within the Kolmogorov cascade.  Since growth is in the unshaded region, the diffusion approximation applies.  See text for further discussion.}
	\label{fig:turbgrowmap}
\end{figure*}

\section{Linear growth with Turbulence}

\label{sec:Turbulent_Background}

In the previous sections, we studied the linear and non-linear behavior of the DSI in the absence of background turbulence, i.e.\ turbulence from sources other than the settling instability itself.  
Such turbulence is relevant because it is expected in realistic discs, and also because it can lift particles away from the midplane, offering opportunities for the DSI after the initial sedimentation phase.  
We thus consider the role of turbulence both for particles settling at $\sim H$ and for cases where the turbulence self-consistently sets the thickness of the dust layer, $H_{\rm d}$. In each case, we show that the turbulence has a strong stabilizing effect on the DSI, by damping growth at shorter wavelengths. 

We characterize the turbulence with the standard $\alpha$ parameter, which sets the turbulent gas viscosity $\nu = \alpha c_{\rm s}H$ \citep{shakura}.  
The diffusion of dust, $D = \nu$, takes the same value, a reasonable approximation for small, well-coupled grains \citep{YoudinLithwik2007}.
The diffusion approximation only holds on scales larger than the large ``outer scale" eddies that dominate the turbulent energy \citep[see e.g.,][]{fan_zhu_1998}. 
Following  \citet[e.g.,][]{Dubrulle1995}, we set the outer scale to
\begin{equation}
l_{\rm eddy} = \sqrt{\alpha}H.
\label{Eq:leddy}
\end{equation}
Thus for wavelengths larger than $l_{\rm eddy}$, we modify the laminar equations of motion (Eqs.\,\ref{Eq:linear_beg}-\ref{Eq:linear_end}) to include turbulent viscosity and dust diffusion as described in Appendix\,\ref{sec:diffusion_appendix}.
For scales smaller than $l_{\rm eddy}$, the turbulence damps modes on the timescale of eddy turnover, at that scale, as described in SH18 and discussed here in Appendix\,\ref{sec:subeddy}.  
There we will show that this small scale damping is strong enough that the long wavelength, i.e.\ diffusive, regime $\lambda > l_{\rm eddy}$ is most relevant.

\subsection{Settling at $H$}

We first consider dust particles settling at one gas scale height from the midplane.  This case was considered by \citetalias{Squire2018} and in the previous sections of this work; it is appropriate when well-stirred particles begin to settle to the midplane. 

Fig.\,\ref{fig:turbgrowmap}  presents two examples of DSI growth rates vs. wavenumbers in the presence of turbulence.  In both, growth only occurs in the diffusive regime, i.e.\ with $k < 2 \pi/l_{\rm eddy}$. 
The leftmost plot adds turbulence with $\alpha = 10^{-6}$ to the case considered in \Fig{fig:fig1}.  Turbulence restricts the growth to longer wavelengths, as expected for diffusion.  In this case, growth is centered on the resonance condition.   
The peak growth rate in this case, $\sigma \simeq 0.009\varOmega_0$,  is only slightly slower than the settling rate of $T_{\rm sett}^{-1} = T_{\rm s} = 0.01$.  Thus very little DSI growth would occur before particles reach lower $z_0$, where growth is slower.

The right plot in \Fig{fig:turbgrowmap} considers  stronger turbulence, $\alpha = 10^{-4}$, where larger values of $\epsilon$ and  $T_{\rm s}$ are chosen to allow DSI growth.  In this case, the peak growth rates are no longer on the resonant curve, but rather occur for smaller $k$.  Specifically, growth is maximized at the largest allowed wavelength $\lambda_z = H$. This result complicates efforts to find (analytically) the maximum level of turbulence that permits DSI growth, as growth away from the resonant condition must also be considered.

\begin{figure*}
	\includegraphics[]{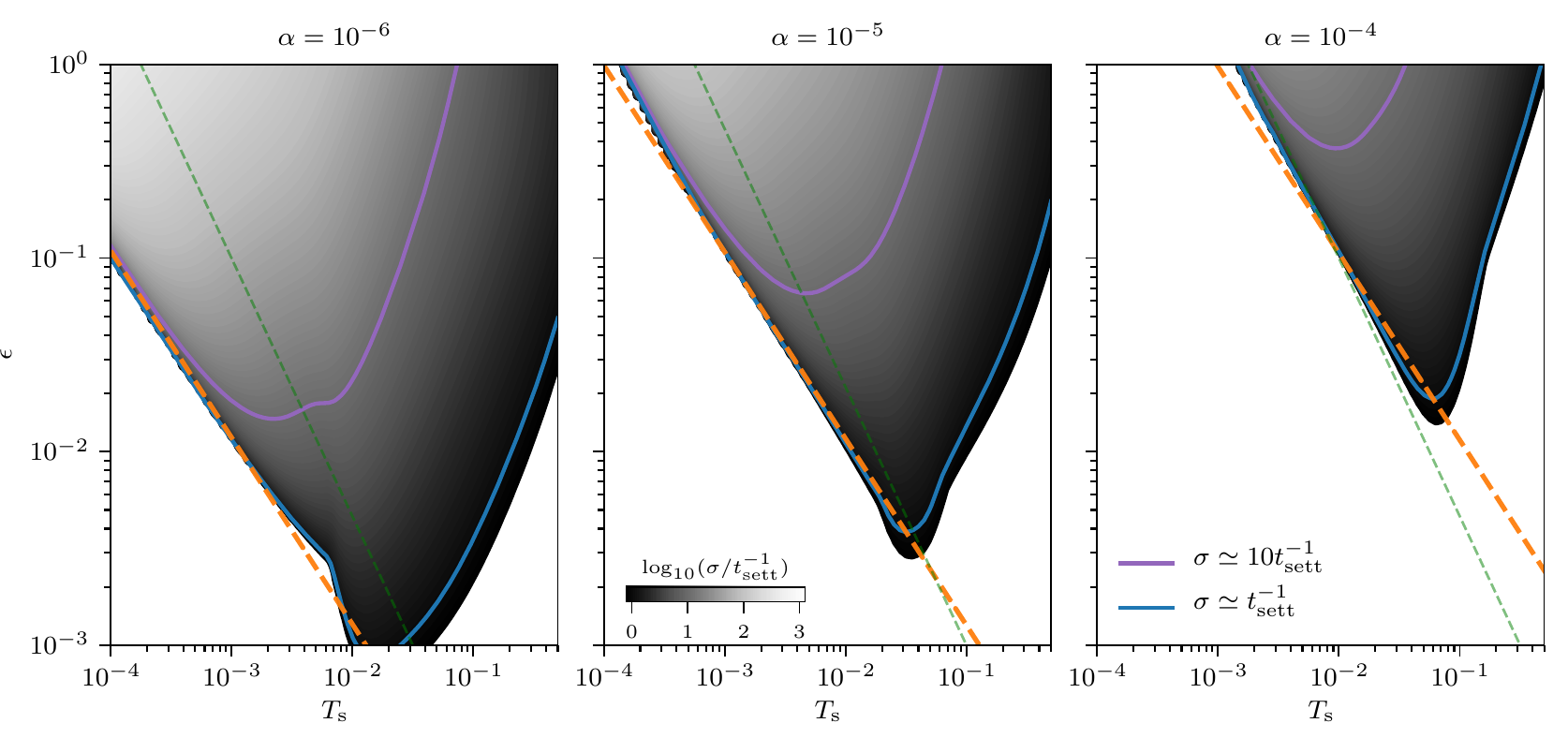} 
	\caption{Colormaps indicate DSI growth rates with turbulence for particles one gas scale height away from the midplane.  Growth rates are normalized by the settling rate, and plotted as a function of the dust-to-gas mass ratio ($\epsilon$) and Stokes number ($T_{\rm s}$).  
	The orange dashed line plots Eq\,\eqref{eq:alpha1}, a fit to the low $T_{\rm s}$ edge of the stability boundary.  
	For reference, the green dashed line shows the stability condition of \citetalias{Squire2018} (see Eq\,\eqref{eq:alphash18}). }
	\label{fig:fig8}
\end{figure*}

Fig.\,\ref{fig:fig8} explores more systematically how turbulence affects the growth of DSI (still at $z_0 = H$), for three different cases: $\alpha=10^{-6},10^{-5}$ and $10^{-4}$.   For each turbulence case, a dense grid of $\epsilon$ and $T_{\rm s}$ values was considered.  For each parameter pair, the fastest growth rate was found over a range of radial and vertical wavenumbers (as in \Fig{fig:fig2} without turbulence).  Specifically $[k_{x,{\rm min}}, k_{x,{\rm max}}] = [k_{z,{\rm min}}, k_{z,{\rm max}}] = 2\pi[1/z_0, 1/l_{\rm eddy}]$, with $z_0 =H$ here.  This upper range of wavenumbers is sufficient because the fastest growth is safely in the diffusive regime, giving no reason to consider $k \gtrsim 2 \pi/l_{\rm eddy}$.  The lower range of wavenumbers is justified as in Section \ref{sec:linear}.  To judge the significance of growth rates, all panels have contours for a growth rate equal to the settling rate (only a small amount of growth) and 10 times the settling rate (more significant growth).

For weak turbulence with $\alpha = 10^{-6}$, significant DSI growth requires $\epsilon \gtrsim 0.02$ and $T_{\rm s} \sim 0.001$ --- $0.01$.  We consider this case in some detail as it is the most favorable that we find.  To judge whether such conditions are likely would require a detailed disc evolution and coagulation model, which we leave to future work.  Nevertheless we note several factors that make DSI growth challenging, even in this favorable low turbulence case and ignoring (for now) that growth does not equal clumping.   The overall abundance of solids at $H$ would need to be above Solar abundances at $H$ implying that very little settling has occurred.  However, to reach these values of $T_s$, significant particle growth would have to occur.  While particle sizes for a given $T_{\rm s}$ depend on highly uncertain gas densities, sub-micron ISM grains should have $T_{\rm s} \lesssim 10^{-5}$ --- $10^{-8}$ from 100 --- 1 AU (i.e. larger $T_{\rm s}$ at larger radii) in a young massive disc \citep{Youdin2010}.  
Particle coagulation by several orders of magnitude in size (more in the inner disc) is unlikely to occur during the early evolution of PPDs. 

We thus describe the following scenario as most favorable to DSI growth.  First, grains grow either (a) with turbulence of $\alpha \gtrsim T_{\rm s}$ so that grains have $H_{\rm d} \sim H$  or (b) grains grow in lower turbulence and are then lofted to $H_{\rm d} \sim H$ by an increase in turbulence to  $\alpha \gtrsim T_{\rm s}$.  Subsequently, in less than one settling time, i.e. in $\lesssim 1/(2\pi T_{\rm s}) \sim 10$ ---$100$ orbital periods, turbulence decreases drastically to $\alpha \ll T_{\rm s}$, allowing DSI growth.  Such a scenario is consistent with  turbulent-quiescent limit cycles \citep{Martin:2011} that could be responsible for FU Orionis outbursts and related variability phenomena \citep{Hartmann2016}. 
However, even in this favorable scenario, DSI growth may not trigger strong clumping, according to the results of Section \ref{sec:NonLinear}, because $\epsilon$ and $T_{\rm s}$ values would not be large enough.

\Fig{fig:fig8} (in the center and right panels) shows that as turbulence increases, the conditions required for DSI growth become more extreme.  
Strongly super-Solar dust-to-gas mass ratios are required to obtain a growth-rate significantly larger than the settling rate.  Specifically, significant growth requires $\epsilon \gtrsim 0.1$ ($0.5$) for $\alpha = 10^{-5}$ ($10^{-4}$, respectively), while the required range of $T_{\rm s}$ increases only modestly.

When we fit the left edge of the colormap of the growth-rate, we get a turbulent stability boundary in \Fig{fig:fig8} of
\begin{align}
\frac{T_{\rm s, min}}{10^{-2}} &\simeq 0.93 \left(\frac{\alpha}{10^{-4}} \right)^{0.99 } \left(\frac{\epsilon}{10^{-1}} \right)^{-1.0},
\label{eq:alpha1}
\end{align} 
for the smallest solids that allow DSI growth.  Including statistical uncertainty, the coefficient and exponents above are, respectively: $  \left( 0.93\pm 0.06 \right), \left(0.992 \pm 0.008\right), \left(-1.0 \pm 0.1\right)$. 
For comparison \citetalias{Squire2018} estimated that the effects of turbulence would be important near\footnote{\citetalias{Squire2018} estimated a range of prefactors, 0.003 -- 0.03.  We take the median of $\sim 0.01$ for simplicity.}
\begin{align}
\frac{T_{\rm s, SH18}}{10^{-2}} &\sim \left(\frac{\alpha}{10^{-4}} \right)^{1/2} \left(\frac{\epsilon}{10^{-1}} \right)^{-3/4}  \, ,
\label{eq:alphash18}
\end{align} 
We plot both our fit and the previous estimate in \Fig{fig:fig8}.  The \citetalias{Squire2018} estimate is reasonably accurate given that different assumptions were made and numerical analysis was not used.   The empirical fit, by design, gives a better criteria over the range of parameters considered.

The growth of DSI with turbulence could be further limited because our most 
unstable modes have $\lambda_z \sim H$, our imposed upper limit across most of the parameter space. At this scale, vertically global effects, specifically stratification, could reduce growth rates, e.g. due to the stabilizing effects of buoyancy.  
Therefore, a global analysis is warranted to examine this effect.

\subsection{Particle Scale Height Set by Turbulence}

\begin{figure*}
	\includegraphics[]{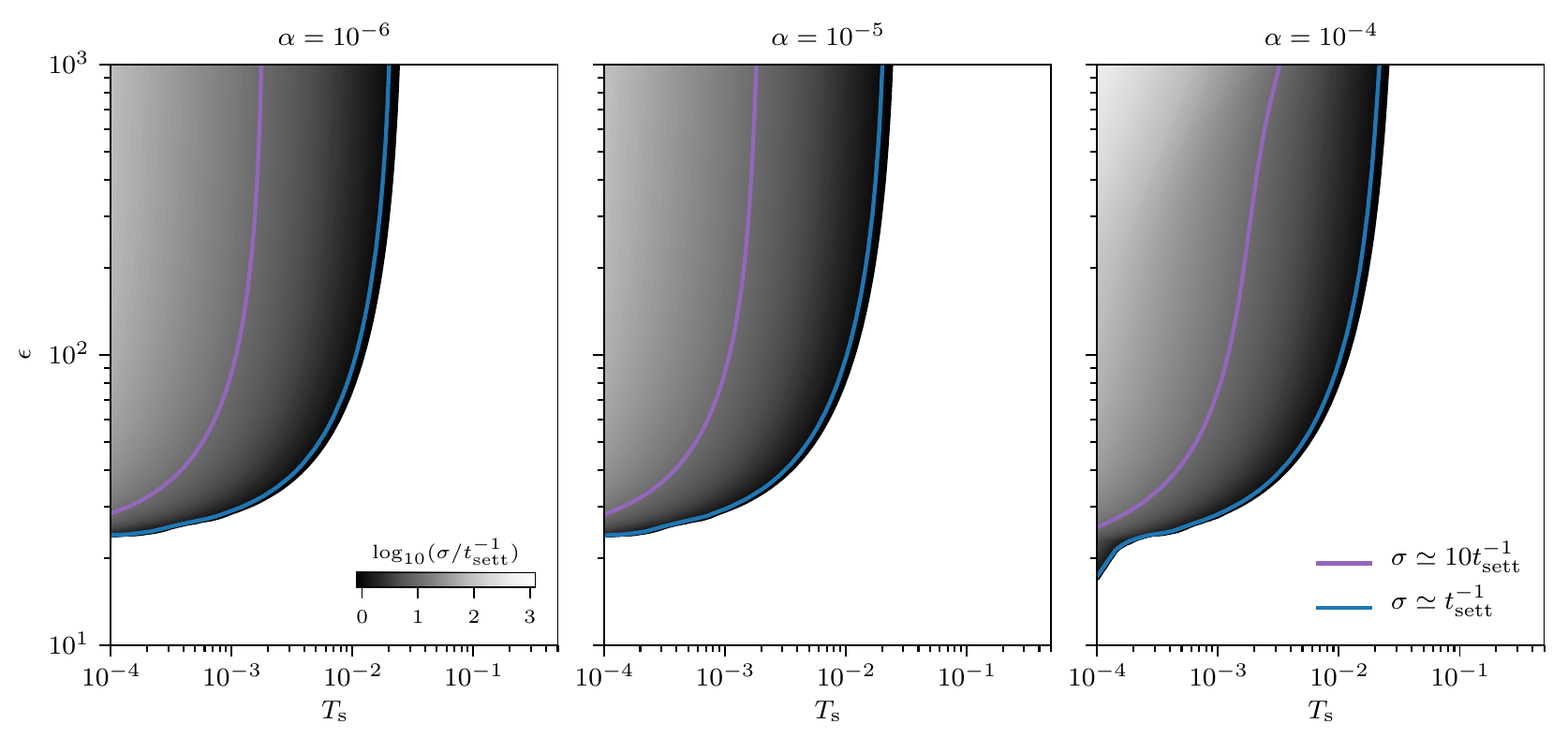} 
	\caption{Same as Fig.\,\ref{fig:fig8} except particles settle from the dust scale height, i.e.\ the equilibrium balance between settling  and turbulent diffusion.  Also the $\epsilon$ values are much larger.   
	In this case, unstable modes only grow for $\epsilon \gtrsim 30$.  Such large values would require some other clumping mechanism to seed the DSI.}
	\label{fig:fig9}
\end{figure*}

Fig.\,\ref{fig:fig9} presents the conditions for DSI growth when the particle scale height is set by the level of turbulence.  In this case, particles settle from  the equilibrium dust scale height, $H_{\rm d} = \sqrt{\alpha/T_{\rm s}}H$ \citep{Dubrulle1995, YoudinLithwik2007}, a result that assumes $\alpha \lesssim T_{\rm s}$, which is indeed required for DSI growth. Contrary to the previous case, particles should persist at $H_{\rm d}$ over much of the lifetime of the disc, i.e.\ apart from sudden changes to $\alpha$ due to accretion outbursts.  This analysis sets $z_0 = H_{\rm d}$, which affects both settling speeds and the allowed wavenumbers, as described above for the $z_0 = H$ case. \footnote{As with the $z_0 = H$ case,  the most unstable radial wavelengths are found to satisfy $\lambda_x < z_0$ so there is no need to consider smaller radial wavenumbers.}

Fig.\,\ref{fig:fig9} shows that when considering settling from smaller $z_0$, very large dust-to-gas mass ratios are needed for DSI growth.
For $T_s = 10^{-4}$, significant growth requires $\epsilon \gtrsim 30$ for all turbulent strengths.  Larger grains (i.e.\ large $T_{\rm s}$) require even more extreme dust-to-gas mass ratios.  This results shows that the DSI  is not relevant in gas rich discs when particles are allowed to settle.  It arises for several physical reasons.  First, as particles settle from smaller $H_{\rm d}$, the settling speeds that drive the DSI are reduced.  Secondly, with reduced $H_{\rm d}$, only smaller vertical wavelengths can grow, but these smaller wavelengths are more readily damped.  Changing the strength of turbulence has little effect because stronger turbulence increases diffusive stabilization while weaker turbulence allows more settling.  Unlike standard midplane SI, increasing $T_{\rm s}$ towards unity does not help DSI growth, again because $H_{\rm d}$ is reduced.

Thus we have shown that even weak turbulence strongly stabilizes the DSI, both when considering settling from $H$ and $H_{\rm d}$.  If dust settles from the equilibrium height $H_{\rm d}$, DSI growth with any turbulence requires extreme $\epsilon$ values.  While settling from $H$ allows more turbulence and lower $\epsilon \sim 0.02$, these settling conditions  only arise briefly after a sharp drop in $\alpha$ following an accretion outburst.  Finally, even in this favorable growth regime, our numerical simulations indicate that the DSI only drives weak particle clumping.

\section{Conclusions}
\label{sec:Discussion}

This paper studies the linear and non-linear behavior of the dust settling instability (DSI) of \citetalias{Squire2018}.  
We conclude that the DSI is unlikely to produce sufficient particle clumping to aid planetesimal formation. We posit that the DSI could produce turbulence that affects the collisional evolution of dust, but only in a fairly specific scenario, e.g. during a period of weak turbulence immediately following an accretion disc outburst that stirs larger dust grains away from the midplane.   We briefly summarize and contextualize these results.

Planetesimal formation via the rapid gravitational collapse of solid particles can only occur if particle densities exceed the Roche density \citep[e.g.,][]{Sekiya1983, Youdin2011GI}, or equivalently if the radius of an overdense clump is within its own Hill Sphere.    
The Roche density exceeds the midplane gas density, by factors $>100$ for standard ``minimum mass" disk models \citep{Hayashi, Chiang2010}.  In more massive disks, the Roche density is closer to the gas density, becoming similar for gravitationally unstable gas disks.  The particle clumping driven by the non-linear evolution of the SI can be strong enough to form planetesimals \citep{Johansen2009b,Simon2017,shaferU2017, LiRixin2019,Nesvorny2019}, even with moderate background turbulence of $\alpha \approx 10^{-3.5}$ \citep{Gole2020}.   
The two-fluid SI is most effective when a significant mass of solids first grows to large sizes with $T_{\rm s} \sim 0.1$ -- $1$ \citep{Yang2017}.

In contrast, the linear analysis of the DSI without turbulence indicates that it might concentrate solids with smaller $T_{\rm s}$. Thus the DSI could in principle aid planetesimal formation, either by triggering direct collapse above the midplane or by seeding the SI \citepalias{Squire2018}.  However, our FARGO3D simulations of the non-linear evolution of the DSI in the laminar case find that particle clumping is weak.  Significant particle clumping was seen only in local simulations with large dust-to-gas ratios ($\epsilon = 0.1$) and large particles ( $T_{\rm s} = 0.1$).  However, the settling time of these particles is faster than the time for significant clumping.  The only simulated case with significant clumping would  not occur a real disk.

Our simulations do indicate that the DSI can drive moderate turbulence $\alpha \sim 10^{-4}$, for near Solar abundances, and values of $T_s$ corresponding to $mm$ grains in the MMSN. Thus several orders of magnitude of grain growth are still required starting from the sub-micron sized dust inherited from the ISM. The impact of DSI turbulence on  disc evolution and grain growth  is an area for future study.

We also explore the linear behavior of the DSI with a dust size distribution. We find that the time and length scales of the instability are comparable  when considering a dust size distribution versus a monodisperse case. Finally, we show that even weak turbulence stabilizes the DSI over the majority of relevant parameter space. 
For particles settling from the equilibrium height $H_{\rm d} \lesssim H$, the system is linearly stable for all expected dust-to-gas mass ratios (above the mid-plane), i.e. $\epsilon \lesssim 1$.  
For particles settling from $H$ the DSI could grow with weak turbulence, $\alpha \sim 10^{-6}$, if there is a significant abundance $\epsilon \sim 0.02$, of intermediate-sized grains with $T_{\rm s} \sim 0.001$ --- $0.01$.  However these favorable growth conditions could only exist briefly, for $\sim 10$ --- 100 orbital periods after a sharp drop in $\alpha$ following an accretion outburst.

\section{Caveats and Future Work}
\label{sec:future}
Our numerical simulations are a useful first step towards understanding which scales and parameters allow the growth of the DSI and the development of strong dust-density enhancements. We propose several promising areas for future study.

Our simulations have yet to fully converge as described in more detail Section \ref{sec:conv}. In 2D, the gas and dust turbulent velocities converge, but the maximum dust densities and non-linear growth rates do not, and in 3D, the saturated turbulent $\alpha$ values and dust concentrations are reduced compared to the equivalent 2D run. Confirmation of our results requires higher-resolution 3D simulations with both two-fluid and particle based codes, as well as different box sizes, boundary conditions, and ultimately self-gravity. Although the effect of stratification on the non-linear phase of the DSI should also be studied in future work, the local approximation employed here is justified for well-coupled grains settling from the gas pressure scale height or below.  

We have studied the linear growth of the DSI in the presence of isotropic background turbulence.  For all relevant cases (i.e. growing modes) we found that the eddy scales were smaller than the growing wavelengths.  Thus our treatment of turbulence as a diffusive process is consistent with standard Reynolds averaging.
Nevertheless the detailed interplay between a turbulent flow and the DSI should be addressed using numerical simulations, especially for  largescale and/or anisotropic turbulence. 
In particular, future work should self-consistently study the DSI in the presence of the Vertical-Shear-Instability (VSI \cite{Nelson2013}). The VSI  induces weak  -- but non-negligible -- turbulence, affecting the settling of solids, as the large-scale eddies lift small dust-grains above the midplane \citep[e.g.,][]{Stoll2016,Flock2017,Schafer2020}, and thus might alter the concentration and distribution of the small grains subject to the DSI.
Noteably, \cite{LinMinKai2019} has shown that  dust feedback delays the growth of the VSI during the settling of particles, but found no evidence of the DSI in simulations. The lack of DSI growth could be attributable to resolution, as we show 1000 cells per $H$ are needed.

While we showed that the growth rates of the DSI converge for particle size distributions, the different saturation timescales for different dust-species leave open the possibility that systems with dust distributions might evolve to a different non-linear state, for example in terms of clumping. 
In particular, future multi-species simulations will shed light on whether the feedback between different dust-species can either induce faster saturation or quench the turbulence and reduce the maximum density concentrations.
Furthermore, the multiple-species DSI dynamics may be of particular interest when interpreting the role of small dust grains in scattered light disc images and SEDs \citep[e.g.,][]{Dullemond2004}.
The action of the DSI in laminar accretion regimes might provide a source for well-mixed dust in disc atmospheres.

\bigskip

\section*{Acknowledgements}
We thank the referee Darryl Seligman for his thorough and comprehensive report, in particular for encouraging us to include a Fourier spectral analysis of the numerical simulations.
We thank Jonathan Squire for helpful discussions which lead to the inclusion of a new simulation to study the convergence with box size.
We thank Martin Pessah for useful discussions and helpful suggestions that improved the clarity of the manuscript. 
We also thanks Philip Hopkins for his valuable comments.
We acknowledge support from Grant 80NSSC19K0639 and useful discussions with members of the TCAN collaboration.  ANY acknowledges support from NASA Astrophysics Theory Grant NNX17AK59G and from NSF grant AST-1616929. Numerical simulations were powered by the El Gato supercomputer supported by the National Science Foundation under Grant No. 1228509. This project has benefited from a collaboration carried out as part of the European Union’s Horizon 2020 research and innovation programme under the Marie Skłodowska-Curie grant agreement No 823823 (DUSTBUSTERS)

\section*{Data availability}

The data underlying this article will be shared on reasonable request to the corresponding author.
All the data was generated with the open-source softwares FARGO3D (available at https://bitbucket.org/fargo3d/public.git, sha:	
9240989), and Multispecies\_si (available at https://bitbucket.org/krappleo/multispecies\_si.git, sha:	
0b945b5).




\bibliographystyle{mnras}

\begin{thebibliography}{}
	\makeatletter
	\relax
	\def\mn@urlcharsother{\let\do\@makeother \do\$\do\&\do\#\do\^\do\_\do\%\do\~}
	\def\mn@doi{\begingroup\mn@urlcharsother \@ifnextchar [ {\mn@doi@}
		{\mn@doi@[]}}
	\def\mn@doi@[#1]#2{\def\@tempa{#1}\ifx\@tempa\@empty \href
		{http://dx.doi.org/#2} {doi:#2}\else \href {http://dx.doi.org/#2} {#1}\fi
		\endgroup}
	\def\mn@eprint#1#2{\mn@eprint@#1:#2::\@nil}
	\def\mn@eprint@arXiv#1{\href {http://arxiv.org/abs/#1} {{\tt arXiv:#1}}}
	\def\mn@eprint@dblp#1{\href {http://dblp.uni-trier.de/rec/bibtex/#1.xml}
		{dblp:#1}}
	\def\mn@eprint@#1:#2:#3:#4\@nil{\def\@tempa {#1}\def\@tempb {#2}\def\@tempc
		{#3}\ifx \@tempc \@empty \let \@tempc \@tempb \let \@tempb \@tempa \fi \ifx
		\@tempb \@empty \def\@tempb {arXiv}\fi \@ifundefined
		{mn@eprint@\@tempb}{\@tempb:\@tempc}{\expandafter \expandafter \csname
			mn@eprint@\@tempb\endcsname \expandafter{\@tempc}}}
	
	\bibitem[\protect\citeauthoryear{{Abod}, {Simon}, {Li}, {Armitage}, {Youdin}
		\& {Kretke}}{{Abod} et~al.}{2019}]{Abod2019}
	{Abod} C.~P.,  {Simon} J.~B.,  {Li} R.,  {Armitage} P.~J.,  {Youdin} A.~N.,
	{Kretke} K.~A.,  2019, \mn@doi [\apj] {10.3847/1538-4357/ab40a3}, \href
	{https://ui.adsabs.harvard.edu/abs/2019ApJ...883..192A} {883, 192}
	
	\bibitem[\protect\citeauthoryear{{Adachi}, {Hayashi}  \& {Nakazawa}}{{Adachi}
		et~al.}{1976}]{Adachi1976}
	{Adachi} I.,  {Hayashi} C.,   {Nakazawa} K.,  1976, Prog.~Theor.~Phys., 56,
	1756
	
	\bibitem[\protect\citeauthoryear{Anderson et~al.,}{Anderson
		et~al.}{1999}]{lapack}
	Anderson E.,  et~al., 1999, {LAPACK} Users' Guide, third edn.
	Society for Industrial and Applied Mathematics, Philadelphia, PA
	
	\bibitem[\protect\citeauthoryear{{Auffinger} \& {Laibe}}{{Auffinger} \&
		{Laibe}}{2018}]{Auffinger2018}
	{Auffinger} J.,  {Laibe} G.,  2018, \mn@doi [\mnras] {10.1093/mnras/stx2395},
	\href {http://adsabs.harvard.edu/abs/2018MNRAS.473..796A} {473, 796}
	
	\bibitem[\protect\citeauthoryear{{Bai} \& {Stone}}{{Bai} \&
		{Stone}}{2010a}]{Bai2010}
	{Bai} X.-N.,  {Stone} J.~M.,  2010a, \mn@doi [\apjs]
	{10.1088/0067-0049/190/2/297}, \href
	{http://adsabs.harvard.edu/abs/2010ApJS..190..297B} {190, 297}
	
	\bibitem[\protect\citeauthoryear{{Bai} \& {Stone}}{{Bai} \&
		{Stone}}{2010b}]{Bai2010ab}
	{Bai} X.-N.,  {Stone} J.~M.,  2010b, \mn@doi [\apj]
	{10.1088/0004-637X/722/2/1437}, \href
	{http://adsabs.harvard.edu/abs/2010ApJ...722.1437B} {722, 1437}
	
	\bibitem[\protect\citeauthoryear{Ben{\'\i}tez-Llambay \&
		Masset}{Ben{\'\i}tez-Llambay \& Masset}{2016}]{Benitez-Llambay2016}
	Ben{\'\i}tez-Llambay P.,  Masset F.~S.,  2016, \mn@doi [\apjs]
	{10.3847/0067-0049/223/1/11}, 223, 11
	
	\bibitem[\protect\citeauthoryear{{Ben{\'{\i}}tez-Llambay}, {Krapp}  \&
		{Pessah}}{{Ben{\'{\i}}tez-Llambay} et~al.}{2019}]{Benitez-Llambay2019}
	{Ben{\'{\i}}tez-Llambay} P.,  {Krapp} L.,   {Pessah} M.~E.,  2019, \mn@doi
	[\apjs] {10.3847/1538-4365/ab0a0e}, \href
	{http://adsabs.harvard.edu/abs/2019ApJS..241...25B} {241, 25}
	
	\bibitem[\protect\citeauthoryear{{Birnstiel}, {Fang}  \&
		{Johansen}}{{Birnstiel} et~al.}{2016}]{Birnstiel2016}
	{Birnstiel} T.,  {Fang} M.,   {Johansen} A.,  2016, \mn@doi [\ssr]
	{10.1007/s11214-016-0256-1}, \href
	{https://ui.adsabs.harvard.edu/abs/2016SSRv..205...41B} {205, 41}
	
	\bibitem[\protect\citeauthoryear{{Blum}}{{Blum}}{2018}]{Blum2018}
	{Blum} J.,  2018, \mn@doi [\ssr] {10.1007/s11214-018-0486-5}, \href
	{https://ui.adsabs.harvard.edu/abs/2018SSRv..214...52B} {214, 52}
	
	\bibitem[\protect\citeauthoryear{Chiang \& Youdin}{Chiang \&
		Youdin}{2010}]{Chiang2010}
	Chiang E.,  Youdin A.,  2010, \mn@doi [Annual Review of Earth and Planetary
	Sciences] {10.1146/annurev-earth-040809-152513}, 38, 493
	
	\bibitem[\protect\citeauthoryear{{D'Alessio}, {Calvet}, {Hartmann},
		{Franco-Hern{\'a}ndez}  \& {Serv{\'\i}n}}{{D'Alessio}
		et~al.}{2006}]{DAlessio2006}
	{D'Alessio} P.,  {Calvet} N.,  {Hartmann} L.,  {Franco-Hern{\'a}ndez} R.,
	{Serv{\'\i}n} H.,  2006, \mn@doi [\apj] {10.1086/498861}, \href
	{https://ui.adsabs.harvard.edu/abs/2006ApJ...638..314D} {638, 314}
	
	\bibitem[\protect\citeauthoryear{{Dipierro}, {Laibe}, {Alexander}  \&
		{Hutchison}}{{Dipierro} et~al.}{2018}]{Dipierro2018}
	{Dipierro} G.,  {Laibe} G.,  {Alexander} R.,   {Hutchison} M.,  2018, \mn@doi
	[\mnras] {10.1093/mnras/sty1701}, \href
	{http://adsabs.harvard.edu/abs/2018MNRAS.479.4187D} {479, 4187}
	
	\bibitem[\protect\citeauthoryear{{Dohnanyi}}{{Dohnanyi}}{1969}]{Dohnanyi1969}
	{Dohnanyi} J.~S.,  1969, \mn@doi [\jgr] {10.1029/JB074i010p02531}, \href
	{http://adsabs.harvard.edu/abs/1969JGR....74.2531D} {74, 2531}
	
	\bibitem[\protect\citeauthoryear{{Dubrulle}, {Morfill}  \&
		{Sterzik}}{{Dubrulle} et~al.}{1995}]{Dubrulle1995}
	{Dubrulle} B.,  {Morfill} G.,   {Sterzik} M.,  1995, \mn@doi [\icarus]
	{10.1006/icar.1995.1058}, \href
	{https://ui.adsabs.harvard.edu/abs/1995Icar..114..237D} {114, 237}
	
	\bibitem[\protect\citeauthoryear{{Dullemond} \& {Dominik}}{{Dullemond} \&
		{Dominik}}{2004}]{Dullemond2004}
	{Dullemond} C.~P.,  {Dominik} C.,  2004, \mn@doi [\aap]
	{10.1051/0004-6361:20040284}, \href
	{https://ui.adsabs.harvard.edu/abs/2004A&A...421.1075D} {421, 1075}
	
	\bibitem[\protect\citeauthoryear{Epstein}{Epstein}{1924}]{Epstein1923}
	Epstein P.~S.,  1924, \mn@doi [Phys. Rev.] {10.1103/PhysRev.23.710}, 23, 710
	
	\bibitem[\protect\citeauthoryear{Fan \& Zhu}{Fan \& Zhu}{1998}]{fan_zhu_1998}
	Fan L.-S.,  Zhu C.,  1998, Principles of Gas-Solid Flows.
	Cambridge Series in Chemical Engineering, Cambridge University Press,
	\mn@doi{10.1017/CBO9780511530142}
	
	\bibitem[\protect\citeauthoryear{{Flock}, {Nelson}, {Turner}, {Bertrang},
		{Carrasco-Gonz{\'a}lez}, {Henning}, {Lyra}  \& {Teague}}{{Flock}
		et~al.}{2017}]{Flock2017}
	{Flock} M.,  {Nelson} R.~P.,  {Turner} N.~J.,  {Bertrang} G. H.~M.,
	{Carrasco-Gonz{\'a}lez} C.,  {Henning} T.,  {Lyra} W.,   {Teague} R.,  2017,
	\mn@doi [\apj] {10.3847/1538-4357/aa943f}, \href
	{https://ui.adsabs.harvard.edu/abs/2017ApJ...850..131F} {850, 131}
	
	\bibitem[\protect\citeauthoryear{{Goldreich} \& {Lynden-Bell}}{{Goldreich} \&
		{Lynden-Bell}}{1965}]{Goldreich1965}
	{Goldreich} P.,  {Lynden-Bell} D.,  1965, \mn@doi [\mnras]
	{10.1093/mnras/130.2.125}, \href
	{https://ui.adsabs.harvard.edu/abs/1965MNRAS.130..125G} {130, 125}
	
	\bibitem[\protect\citeauthoryear{{Goldreich} \& {Ward}}{{Goldreich} \&
		{Ward}}{1973}]{Goldreich1973}
	{Goldreich} P.,  {Ward} W.~R.,  1973, \mn@doi [\apj] {10.1086/152291}, \href
	{https://ui.adsabs.harvard.edu/abs/1973ApJ...183.1051G} {183, 1051}
	
	\bibitem[\protect\citeauthoryear{{Gole}, {Simon}, {Li}, {Youdin}  \&
		{Armitage}}{{Gole} et~al.}{2020}]{Gole2020}
	{Gole} D.~A.,  {Simon} J.~B.,  {Li} R.,  {Youdin} A.~N.,   {Armitage} P.~J.,
	2020, arXiv e-prints, \href
	{https://ui.adsabs.harvard.edu/abs/2020arXiv200110000G} {p. arXiv:2001.10000}
	
	\bibitem[\protect\citeauthoryear{{G{\'o}mez} \& {Ostriker}}{{G{\'o}mez} \&
		{Ostriker}}{2005}]{Gomez2005}
	{G{\'o}mez} G.~C.,  {Ostriker} E.~C.,  2005, \mn@doi [\apj] {10.1086/432086},
	\href {https://ui.adsabs.harvard.edu/abs/2005ApJ...630.1093G} {630, 1093}
	
	\bibitem[\protect\citeauthoryear{{Goodman} \& {Pindor}}{{Goodman} \&
		{Pindor}}{2000}]{Goodman2000}
	{Goodman} J.,  {Pindor} B.,  2000, Icarus, \href
	{http://adsabs.harvard.edu/cgi-bin/nph-bib_query?bibcode=2000Icar..148..537G&db_key=AST}
	{148, 537}
	
	\bibitem[\protect\citeauthoryear{{Hartmann}, {Herczeg}  \& {Calvet}}{{Hartmann}
		et~al.}{2016}]{Hartmann2016}
	{Hartmann} L.,  {Herczeg} G.,   {Calvet} N.,  2016, \araa, 54, 135
	
	\bibitem[\protect\citeauthoryear{Hayashi}{Hayashi}{1981}]{Hayashi}
	Hayashi C.,  1981, \mn@doi [Progress of Theoretical Physics Supplement]
	{10.1143/PTPS.70.35}, 70, 35
	
	\bibitem[\protect\citeauthoryear{{Hopkins}, {Squire}  \& {Seligman}}{{Hopkins}
		et~al.}{2020}]{Hopkins2019}
	{Hopkins} P.~F.,  {Squire} J.,   {Seligman} D.,  2020, \mn@doi [\mnras]
	{10.1093/mnras/staa1046}, \href
	{https://ui.adsabs.harvard.edu/abs/2020MNRAS.tmp.1235H} {}
	
	\bibitem[\protect\citeauthoryear{{Johansen} \& {Youdin}}{{Johansen} \&
		{Youdin}}{2007}]{Johansen&Youdin2007}
	{Johansen} A.,  {Youdin} A.,  2007, \mn@doi [\apj] {10.1086/516730}, \href
	{http://adsabs.harvard.edu/abs/2007ApJ...662..627J} {662, 627}
	
	\bibitem[\protect\citeauthoryear{{Johansen}, {Oishi}, {Mac Low}, {Klahr},
		{Henning}  \& {Youdin}}{{Johansen} et~al.}{2007}]{Johansen2007}
	{Johansen} A.,  {Oishi} J.~S.,  {Mac Low} M.-M.,  {Klahr} H.,  {Henning} T.,
	{Youdin} A.,  2007, \mn@doi [\nat] {10.1038/nature06086}, \href
	{https://ui.adsabs.harvard.edu/abs/2007Natur.448.1022J} {448, 1022}
	
	\bibitem[\protect\citeauthoryear{{Johansen}, {Youdin}  \& {Mac Low}}{{Johansen}
		et~al.}{2009}]{Johansen2009b}
	{Johansen} A.,  {Youdin} A.,   {Mac Low} M.-M.,  2009, \mn@doi [\apjl]
	{10.1088/0004-637X/704/2/L75}, \href
	{https://ui.adsabs.harvard.edu/abs/2009ApJ...704L..75J} {704, L75}
	
	\bibitem[\protect\citeauthoryear{{Krapp}, {Ben{\'\i}tez-Llambay}, {Gressel}  \&
		{Pessah}}{{Krapp} et~al.}{2019}]{Krapp2019}
	{Krapp} L.,  {Ben{\'\i}tez-Llambay} P.,  {Gressel} O.,   {Pessah} M.~E.,  2019,
	\mn@doi [\apj] {10.3847/2041-8213/ab2596}, \href
	{https://ui.adsabs.harvard.edu/abs/2019ApJ...878L..30K} {878, L30}
	
	\bibitem[\protect\citeauthoryear{{Krijt} \& {Ciesla}}{{Krijt} \&
		{Ciesla}}{2016}]{Krijt2016}
	{Krijt} S.,  {Ciesla} F.~J.,  2016, \mn@doi [\apj]
	{10.3847/0004-637X/822/2/111}, \href
	{https://ui.adsabs.harvard.edu/abs/2016ApJ...822..111K} {822, 111}
	
	\bibitem[\protect\citeauthoryear{{Lambrechts}, {Johansen}, {Capelo}, {Blum}  \&
		{Bodenschatz}}{{Lambrechts} et~al.}{2016}]{Lambrechts2016}
	{Lambrechts} M.,  {Johansen} A.,  {Capelo} H.~L.,  {Blum} J.,   {Bodenschatz}
	E.,  2016, \mn@doi [\aap] {10.1051/0004-6361/201526272}, \href
	{https://ui.adsabs.harvard.edu/abs/2016A&A...591A.133L} {591, A133}
	
	\bibitem[\protect\citeauthoryear{{Latter} \& {Papaloizou}}{{Latter} \&
		{Papaloizou}}{2017}]{Latter2017}
	{Latter} H.~N.,  {Papaloizou} J.,  2017, \mn@doi [\mnras]
	{10.1093/mnras/stx2038}, \href
	{https://ui.adsabs.harvard.edu/abs/2017MNRAS.472.1432L} {472, 1432}
	
	\bibitem[\protect\citeauthoryear{{Li}, {Youdin}  \& {Simon}}{{Li}
		et~al.}{2019}]{LiRixin2019}
	{Li} R.,  {Youdin} A.~N.,   {Simon} J.~B.,  2019, \mn@doi [\apj]
	{10.3847/1538-4357/ab480d}, \href
	{https://ui.adsabs.harvard.edu/abs/2019ApJ...885...69L} {885, 69}
	
	\bibitem[\protect\citeauthoryear{{Lin}}{{Lin}}{2019}]{LinMinKai2019}
	{Lin} M.-K.,  2019, \mn@doi [\mnras] {10.1093/mnras/stz701}, \href
	{https://ui.adsabs.harvard.edu/abs/2019MNRAS.485.5221L} {485, 5221}
	
	\bibitem[\protect\citeauthoryear{{Martin} \& {Lubow}}{{Martin} \&
		{Lubow}}{2011}]{Martin:2011}
	{Martin} R.~G.,  {Lubow} S.~H.,  2011, \mn@doi [\apjl]
	{10.1088/2041-8205/740/1/L6}, \href
	{http://adsabs.harvard.edu/abs/2011ApJ...740L...6M} {740, L6}
	
	\bibitem[\protect\citeauthoryear{{Masset}}{{Masset}}{2000}]{Masset2000}
	{Masset} F.,  2000, \mn@doi [\aaps] {10.1051/aas:2000116}, \href
	{http://adsabs.harvard.edu/abs/2000A5M26AS..141..16} {141, 165}
		
	\bibitem[\protect\citeauthoryear{{Moseley}, {Squire}  \& {Hopkins}}{{Moseley}
			et~al.}{2019}]{Moseley2019}
		{Moseley} E.~R.,  {Squire} J.,   {Hopkins} P.~F.,  2019, \mn@doi [\mnras]
		{10.1093/mnras/stz2128}, \href
		{https://ui.adsabs.harvard.edu/abs/2019MNRAS.489..325M} {489, 325}
		
		\bibitem[\protect\citeauthoryear{{Nakagawa}, {Sekiya}  \& {Hayashi}}{{Nakagawa}
			et~al.}{1986}]{Nakagawa1986}
		{Nakagawa} Y.,  {Sekiya} M.,   {Hayashi} C.,  1986, \mn@doi [\icarus]
		{10.1016/0019-1035(86)90121-1}, \href
		{http://adsabs.harvard.edu/abs/1986Icar...67..375N} {67, 375}
		
		\bibitem[\protect\citeauthoryear{Nelson, Gressel  \& Umurhan}{Nelson
			et~al.}{2013}]{Nelson2013}
		Nelson R.~P.,  Gressel O.,   Umurhan O.~M.,  2013, \mn@doi [Monthly Notices of
		the Royal Astronomical Society] {10.1093/mnras/stt1475}, 435, 2610
		
		\bibitem[\protect\citeauthoryear{{Nesvorn{\'y}}, {Li}, {Youdin}, {Simon}  \&
			{Grundy}}{{Nesvorn{\'y}} et~al.}{2019}]{Nesvorny2019}
		{Nesvorn{\'y}} D.,  {Li} R.,  {Youdin} A.~N.,  {Simon} J.~B.,   {Grundy} W.~M.,
		2019, \mn@doi [Nature Astronomy] {10.1038/s41550-019-0806-z}, \href
		{https://ui.adsabs.harvard.edu/abs/2019NatAs...3..808N} {3, 808}
		
		\bibitem[\protect\citeauthoryear{{Ormel} \& {Cuzzi}}{{Ormel} \&
			{Cuzzi}}{2007}]{Ormel2007}
		{Ormel} C.~W.,  {Cuzzi} J.~N.,  2007, \mn@doi [\aap]
		{10.1051/0004-6361:20066899}, \href
		{https://ui.adsabs.harvard.edu/abs/2007A&A...466..413O} {466, 413}
		
		\bibitem[\protect\citeauthoryear{{Pinilla} \& {Youdin}}{{Pinilla} \&
			{Youdin}}{2017}]{Pinilla2017}
		{Pinilla} P.,  {Youdin} A.,  2017, in {Pessah} M.,  {Gressel} O.,  eds,
		Astrophysics and Space Science Library Vol. 445, Astrophysics and Space
		Science Library. p.~91, \mn@doi{10.1007/978-3-319-60609-5_4}
		
		\bibitem[\protect\citeauthoryear{{Rettig}, {Brittain}, {Simon}, {Gibb},
			{Balsara}, {Tilley}  \& {Kulesa}}{{Rettig} et~al.}{2006}]{Rettig2006}
		{Rettig} T.,  {Brittain} S.,  {Simon} T.,  {Gibb} E.,  {Balsara} D.~S.,
		{Tilley} D.~A.,   {Kulesa} C.,  2006, \mn@doi [\apj] {10.1086/504703}, \href
		{https://ui.adsabs.harvard.edu/abs/2006ApJ...646..342R} {646, 342}
		
		\bibitem[\protect\citeauthoryear{{Safronov}}{{Safronov}}{1969}]{saf69}
		{Safronov} V.~S.,  1969, {Evoliutsiia doplanetnogo oblaka.}.
		Moscow: Nakua
		
		\bibitem[\protect\citeauthoryear{{Sch{\"a}fer}, {Yang}  \&
			{Johansen}}{{Sch{\"a}fer} et~al.}{2017}]{shaferU2017}
		{Sch{\"a}fer} U.,  {Yang} C.-C.,   {Johansen} A.,  2017, \mn@doi [\aap]
		{10.1051/0004-6361/201629561}, \href
		{https://ui.adsabs.harvard.edu/abs/2017A&A...597A..69S} {597, A69}
		
		\bibitem[\protect\citeauthoryear{{Sch{\"a}fer}, {Johansen}  \&
			{Banerjee}}{{Sch{\"a}fer} et~al.}{2020}]{Schafer2020}
		{Sch{\"a}fer} U.,  {Johansen} A.,   {Banerjee} R.,  2020, \mn@doi [\aap]
		{10.1051/0004-6361/201937371}, \href
		{https://ui.adsabs.harvard.edu/abs/2020A&A...635A.190S} {635, A190}
		
		\bibitem[\protect\citeauthoryear{Sekiya}{Sekiya}{1983}]{Sekiya1983}
		Sekiya M.,  1983, \mn@doi [Progress of Theoretical Physics]
		{10.1143/PTP.69.1116}, 69, 1116
		
		\bibitem[\protect\citeauthoryear{{Seligman}, {Hopkins}  \& {Squire}}{{Seligman}
			et~al.}{2019}]{Seligman2019}
		{Seligman} D.,  {Hopkins} P.~F.,   {Squire} J.,  2019, \mn@doi [\mnras]
		{10.1093/mnras/stz666}, \href
		{https://ui.adsabs.harvard.edu/abs/2019MNRAS.485.3991S} {485, 3991}
		
		\bibitem[\protect\citeauthoryear{{Shakura} \& {Sunyaev}}{{Shakura} \&
			{Sunyaev}}{1973}]{shakura}
		{Shakura} N.~I.,  {Sunyaev} R.~A.,  1973, \aap, \href
		{http://adsabs.harvard.edu/abs/1973A%26A....24..337S} {24, 337}
			
			\bibitem[\protect\citeauthoryear{{Simon}, {Armitage}, {Youdin}  \&
				{Li}}{{Simon} et~al.}{2017}]{Simon2017}
			{Simon} J.~B.,  {Armitage} P.~J.,  {Youdin} A.~N.,   {Li} R.,  2017, \mn@doi
			[\apjl] {10.3847/2041-8213/aa8c79}, \href
			{http://adsabs.harvard.edu/abs/2017ApJ...847L..12S} {847, L12}
			
			\bibitem[\protect\citeauthoryear{{Squire} \& {Hopkins}}{{Squire} \&
				{Hopkins}}{2018a}]{Squire2018}
			{Squire} J.,  {Hopkins} P.~F.,  2018a, \mn@doi [\mnras] {10.1093/mnras/sty854},
			\href {http://adsabs.harvard.edu/abs/2018MNRAS.477.5011S} {477, 5011}
			
			\bibitem[\protect\citeauthoryear{{Squire} \& {Hopkins}}{{Squire} \&
				{Hopkins}}{2018b}]{Squire2018letter}
			{Squire} J.,  {Hopkins} P.~F.,  2018b, \mn@doi [\apjl]
			{10.3847/2041-8213/aab54d}, \href
			{http://adsabs.harvard.edu/abs/2018ApJ...856L..15S} {856, L15}
			
			\bibitem[\protect\citeauthoryear{{Stoll} \& {Kley}}{{Stoll} \&
				{Kley}}{2016}]{Stoll2016}
			{Stoll} M. H.~R.,  {Kley} W.,  2016, \mn@doi [\aap]
			{10.1051/0004-6361/201527716}, \href
			{https://ui.adsabs.harvard.edu/abs/2016A&A...594A..57S} {594, A57}
			
			\bibitem[\protect\citeauthoryear{{Takeuchi} \& {Lin}}{{Takeuchi} \&
				{Lin}}{2002}]{Takeuchi2002}
			{Takeuchi} T.,  {Lin} D.~N.~C.,  2002, \mn@doi [\apj] {10.1086/344437}, \href
			{http://adsabs.harvard.edu/abs/2002ApJ...581.1344T} {581, 1344}
			
			\bibitem[\protect\citeauthoryear{{Tominaga}, {Takahashi}  \&
				{Inutsuka}}{{Tominaga} et~al.}{2019}]{Tominaga2019}
			{Tominaga} R.~T.,  {Takahashi} S.~Z.,   {Inutsuka} S.-i.,  2019, \mn@doi [\apj]
			{10.3847/1538-4357/ab25ea}, \href
			{https://ui.adsabs.harvard.edu/abs/2019ApJ...881...53T} {881, 53}
			
			\bibitem[\protect\citeauthoryear{Walt, Colbert  \& Varoquaux}{Walt
				et~al.}{2011}]{Walt2011}
			Walt S. v.~d.,  Colbert S.~C.,   Varoquaux G.,  2011, \mn@doi [Computing in
			Science and Engg.] {10.1109/MCSE.2011.37}, 13, 22
			
			\bibitem[\protect\citeauthoryear{{Weidenschilling}}{{Weidenschilling}}{1980}]{Weidenschilling1980}
			{Weidenschilling} S.~J.,  1980, \mn@doi [\icarus]
			{10.1016/0019-1035(80)90064-0}, \href
			{https://ui.adsabs.harvard.edu/abs/1980Icar...44..172W} {44, 172}
			
			\bibitem[\protect\citeauthoryear{Weidenschilling}{Weidenschilling}{1984}]{WEIDENSCHILLING1984}
			Weidenschilling S.~J.,  1984, \mn@doi [Icarus]
			{https://doi.org/10.1016/0019-1035(84)90164-7}, 60, 553
			
			\bibitem[\protect\citeauthoryear{{Whipple}}{{Whipple}}{1972}]{Whipple1972}
			{Whipple} F.~L.,  1972, in {Elvius} A.,  ed., From Plasma to Planet. p.~211
			
			\bibitem[\protect\citeauthoryear{{Yang}, {Johansen}  \& {Carrera}}{{Yang}
				et~al.}{2017}]{Yang2017}
			{Yang} C.~C.,  {Johansen} A.,   {Carrera} D.,  2017, \mn@doi [\aap]
			{10.1051/0004-6361/201630106}, \href
			{https://ui.adsabs.harvard.edu/abs/2017A&A...606A..80Y} {606, A80}
			
			\bibitem[\protect\citeauthoryear{{Youdin}}{{Youdin}}{2010}]{Youdin2010}
			{Youdin} A.~N.,  2010, in {T.~Montmerle, D.~Ehrenreich, \& A.-M.~Lagrange} ed.,
			EAS Publications Series Vol. 41, EAS Publications Series. pp 187--207
			
			\bibitem[\protect\citeauthoryear{Youdin}{Youdin}{2011}]{Youdin2011GI}
			Youdin A.~N.,  2011, \mn@doi [\apj] {10.1088/0004-637x/731/2/99}, 731, 99
			
			\bibitem[\protect\citeauthoryear{Youdin \& Goodman}{Youdin \&
				Goodman}{2005}]{Youdin2005}
			Youdin A.~N.,  Goodman J.,  2005, \mn@doi [\apj] {10.1086/426895}, 620, 459
			
			\bibitem[\protect\citeauthoryear{Youdin \& Johansen}{Youdin \&
				Johansen}{2007}]{Youdin2007}
			Youdin A.,  Johansen A.,  2007, \mn@doi [\apj] {10.1086/516729}, 662, 613
			
			\bibitem[\protect\citeauthoryear{{Youdin} \& {Lithwick}}{{Youdin} \&
				{Lithwick}}{2007}]{YoudinLithwik2007}
			{Youdin} A.~N.,  {Lithwick} Y.,  2007, \mn@doi [\icarus]
			{10.1016/j.icarus.2007.07.012}, \href
			{https://ui.adsabs.harvard.edu/abs/2007Icar..192..588Y} {192, 588}
			
			\bibitem[\protect\citeauthoryear{{Youdin} \& {Shu}}{{Youdin} \&
				{Shu}}{2002}]{Youdin2002}
			{Youdin} A.~N.,  {Shu} F.~H.,  2002, \mn@doi [\apj] {10.1086/343109}, \href
			{https://ui.adsabs.harvard.edu/abs/2002ApJ...580..494Y} {580, 494}
			
			\bibitem[\protect\citeauthoryear{{Zhuravlev}}{{Zhuravlev}}{2019}]{Zhuravlev2019}
			{Zhuravlev} V.~V.,  2019, \mn@doi [\mnras] {10.1093/mnras/stz2390}, \href
			{https://ui.adsabs.harvard.edu/abs/2019MNRAS.489.3850Z} {489, 3850}
			
			\bibitem[\protect\citeauthoryear{{Zhuravlev}}{{Zhuravlev}}{2020}]{Zhuravlev2020}
			{Zhuravlev} V.~V.,  2020, \mn@doi [\mnras] {10.1093/mnras/staa805}, \href
			{https://ui.adsabs.harvard.edu/abs/2020MNRAS.494.1395Z} {494, 1395}
			
			\makeatother
\end{thebibliography}

\appendix
\section{Linearized Equations}
\label{sec:linearized}

In this Section we present the linearized  steady-state solution of the system of Eqs.\,\eqref{Eq:set_begin}-\eqref{Eq:set_end} in addition to the normalized linear system that we solve in Sections\,\ref{sec:linear} and \ref{sec:Turbulent_Background}.
We start by decomposing the fluid variables into an axisymmetric perturbation with complex amplitude $\delta f$, and a constant background steady-state\footnote{The background steady-state is calculated relative to the shear velocity ${\bf v}_s = q\varOmega_0x{\bf e}_y$.}, $f^{0}$, as follows
\begin{equation}
f = f^0 + \delta f e^{i{\bf k} \cdot {\bf x} - i\gamma t }
\end{equation} 
with ${\bf k} = (k_x,0,k_z)$ the wavevector and $\gamma$ the mode eigenvalue.
Assuming constant background densities for the gas and dust j$th$-species $\rho^0_{\rm g}$ and $\rho^0_{j}=\epsilon_j\rho^0_{\rm g0}$, respectively, explicit analytical solutions for the steady-state radial and azimuthal velocities can be found for an arbitrary number of dust-species \citep[see Section 3.5.2][]{Benitez-Llambay2019}. 
These solutions reduce to those obtained by \cite{Nakagawa1986} when having a unique dust-species. 
For the purpose of this work, we present the solutions obtained by \cite{Benitez-Llambay2019} assuming Keplerian shear. 
Defining 
\begin{equation}
\psi =\left( \sum_{k=1}^{N} \frac{\epsilon_k T_{{\rm s}k}}{1+T^2_{{\rm s}k}}\right)^2 + \left(1 + \sum_{k=1}^{N} \frac{\epsilon_k}{1+T^2_{{\rm s}k}} \right)^2,
\end{equation}

the steady-state gas velocities normalized by $\eta v_K$ read as
\begin{align}
\label{Eq:steady_state_gas}
\tilde{v}^0_{{\rm g}x} &= 2\psi^{-1}\sum^{N}_{k=1} \frac{\epsilon_k T_{{\rm s}k}}{1+T^2_{{\rm s}k}}, \\
\tilde{v}^0_{{\rm g}y} &= -\psi^{-1} \left(1+\sum_{k=1}^{N} \frac{\epsilon_k }{1+T^2_{{\rm s}k}}\right), 
\end{align} 
while for the $j$th dust-species we have
\begin{align}
\tilde{v}^0_{jx} &= \frac{1}{1 + T^2_{{\rm s}j}} \left( \tilde{v}^0_{{\rm g}x} + 2T_{{\rm s}j}\tilde{v}^0_{{\rm g}y}\right), \\
\tilde{v}_{jy} &= \frac{1}{1 + T^2_{{\rm s}j}} \left(  \tilde{v}^0_{{\rm g}y} - \frac{1}{2}T_{{\rm}j}\tilde{v}^0_{{\rm g}x} \right).
\label{Eq:steady_state_dust}
\end{align}
Thus, the normalized steady-state radial and azimuthal velocities are only a function of the dust-to-gas mass ratio and the Stokes number.
For the vertical velocity, we consider a coordinate system where the gas is at rest while the dust-species is settling from $z=H$ below the mid-plane.
Thus, the gas and dust vertical velocities are
\begin{equation}
\tilde{v}^0_{{\rm g}z} = 0\,\quad \textrm{and}\quad \tilde{v}^0_{j{z}} =  \zeta T_{{\rm s}j}\,,
\end{equation}
only valid for dust-species at the terminal velocity, i.e. species with $T_{\rm s}\ll 1$.
The control parameter $\zeta = z_0/(\eta r_0)$ was introduced in Section\,\ref{sec:Equations}.
In the single dust species case (where we use the subscript ``d" to replace the numerical dust species label) the relative drift velocity that enters the RDI theory is:
\begin{align}\label{eq:eqdrift}
{\bf w} &= \eta v_K (\tilde{\bf{v}}^0_{\rm d} - \tilde{\bf{v}}^0_{\rm g})\, .
\end{align} 
Adding dust species introduces multiple drift speeds relative to the gas.

After replacing the gas and dust-species densities and velocities by $f$ in Eqs.\,\eqref{Eq:set_begin}-\eqref{Eq:set_end} ($f^0$ has to be replaced by the corresponding steady-state solution described above), the eigenvalue $\gamma$ and its associated eigenvector complex amplitude, $\delta f$, is found by solving the linear, dimensionless, and normalized system of equations:
\begin{align}
i \tilde{{\bf k}} \cdot \delta \tilde{{\bf v}}_{\rm g} &= i \tilde{\gamma}_{{\rm g}} \delta \tilde{\rho}_{\rm g} \label{Eq:linear_beg}\\
i\epsilon_j \tilde{{\bf k}} \cdot \delta \tilde{{\bf v}}_{j}   &=  i \tilde{\gamma}_j \delta \tilde{\rho}_{j} \\
i\tilde{\bf k} \delta \tilde{\rho}_{\rm g} \Pi^{-2} + 2 {\bf e}_z \times \delta \tilde {\bf v}_{\rm g} - \frac{3}{2}\delta \tilde{v}_{{\rm g}x}{\bf e}_y +  \delta \tilde{\bf F}_{\rm g}   &= i \tilde{\gamma}_{{\rm g}}  \delta \tilde{{\bf v}}_{\rm g} \label{Eq:lin_mom}\\
2 {\bf e}_z \times \delta \tilde {\bf v}_{j} - \frac{3}{2}\delta \tilde{v}_{{j}x}{\bf e}_y +  \delta \tilde{\bf F}_{j}   &= i \tilde{\gamma}_j \delta \tilde{{\bf v}}_{j}
\label{Eq:linear_end}
\end{align}
for $j=1 \dotsc N$ and where $\tilde{\gamma}_{\rm g} = \gamma/\varOmega_0 - \tilde{{\bf k}} \cdot \tilde{\bf v}^0_{{\rm g}}$ and $\tilde{\gamma}_{j} = \gamma/\varOmega_0 - \tilde{{\bf k}} \cdot \tilde{\bf v}^0_{j}$.
The normalized eigenvalue and wavevector correspond to $\gamma/\varOmega_0$ and $\tilde{{\bf k}}={\bf k}\eta r_0$, respectively. 
The densities and velocities are normalized by $\rho^0_{\rm g}$ and $\eta v_K$, respectively.
In addition, the perturbed drag specific forces follow from:
\begin{align}
\delta \tilde{\bf F}_{\rm g} &= \sum_{k=1}^{N} \frac{\epsilon_k}{T_{{\rm s}k}} \left( \delta \tilde{ {\bf v}}_{\rm g} - \delta \tilde{ {\bf v}}_{k} \right)  \nonumber \\
&+ \frac{1}{T_{{\rm s}k}} \left( \tilde{ {\bf v}}^0_{\rm g} - \tilde{ {\bf v}}^0_{k} \right) \left(\delta \tilde{\rho}_k - \epsilon_k\delta\tilde{\rho}_{\rm g}\right), \\
\delta \tilde{\bf F}_{j} &= \frac{1}{T_{{\rm s}j}} \left( \delta \tilde{ {\bf v}}_{j} - \delta \tilde{ {\bf v}}_{\rm g} \right),
\end{align}
where we neglect the perturbations to the Stokes number as appropriate for a linear drag law (i.e. Epstein's Law) and for nearly incompressible motions. 
Thus, the particle size, $a$, is strictly proportional to the Stokes number $ T_{\rm s} \simeq a\rho_p /(H\rho^0_{\rm g})$,  assuming that all particles are spherical with radius $a$ and an intrinsic density $\rho_p$ \citep[see e.g.,][]{Takeuchi2002}.
It is important to stress that any velocity or density can be obtained by taking the real part of its corresponding complex analog $f$.
Furthermore, because a given pair of wavelengths $\lambda_x=2\pi/k_x\,,\lambda_z=2\pi/k_z$ may admit several unstable solutions, we define the growth rate, $\sigma$, of the instability as the maximum imaginary part of all the obtained $\gamma$-values.

\bigskip  

In this work we solve the eigenvalue problem in a domain spanned between $[ k_{\rm min}, k_{\rm max}]$, where $k_{\rm min}$ and $k_{\rm max}$ are specified in each section. To find the solutions we use the public available multi-species linear solver with parallel capabilities\footnote{krappleo@bitbucket.org/krappleo/multispecies\_si.git}. The core of the solver use the function \verb|eig| of \verb|NumPy| \citep{Walt2011}, which uses LAPACK routines for complex non-symmetric matrices \citep{lapack}.

\section{Diffusion Approximation}
\label{sec:diffusion_appendix}
In Section\,\ref{sec:Turbulent_Background} we consider the effect of particle diffusion and turbulent viscosity. 
At scales larger than $l_{\rm eddy}=\sqrt{\alpha}H$, we adopt the diffusion approximation and include two additional terms in Eqs\,\eqref{Eq:linear_beg}-\eqref{Eq:linear_end}. 
The inclusion of gas viscosity modifies the steady-state solutions described in Appendix\,\ref{sec:linearized} \citep[see e.g,][]{Dipierro2018}, however, this modification can be safely ignored for the parameters studied in Section\,\ref{sec:Turbulent_Background}. 
To better understand why these terms are negligible, recall that the linear growth of the DSI strongly depends on the vertical and radial drift velocities. In a standard viscous accretion model, the radial accretion flow is $v_{\rm visc} \sim \alpha (H/r_0)^2 v_K$ giving a radial drift of $ \tilde{v}^0_{{\rm d}x} - \tilde{v}^0_{{\rm g}x} \sim  \frac{-2 T_{\rm s}}{(1+\epsilon)^2 + T^2_{\rm s}} \left( 1+\epsilon - T_{\rm s}\alpha \right)$. 
Thus the correction, which is $\mathcal{O}(T_{\rm s}\alpha)$, can be neglected in our analysis in Section\,\ref{sec:Turbulent_Background}, as $T_{\rm s}\alpha \ll 1+\epsilon$ for all cases.
Furthermore, \cite{Tominaga2019} has proposed a revised treatment of dust diffusion that enforces momentum conservation.
We neglect these terms to avoid the viscous instabilities that they can introduce. 
Thus, the dust continuity equation is modified including the particle-diffusion term 
\begin{equation}
\partial_t \rho_{j} + \nabla \cdot\left( \rho_{j} \mathbf{v}_{j}\right) = \nabla \cdot \left(\rho_{\rm g} D_j \nabla \left( \frac{\rho_j}{\rho_{\rm g}}\right)\right)\,,
\end{equation}
which in linearized and non-dimensional form becomes
\begin{equation}
i\epsilon_j \left(\tilde{{\bf k}} \cdot \delta \tilde{{\bf v}}_{j}  + i\tilde{D}_j \tilde{k}^2 \delta \tilde{\rho}_{\rm g} \right) = i\left(\tilde{\gamma}_j + i\tilde{D}_j \tilde{k}^2\right) \delta \tilde{\rho}_{j}\,,
\label{Eq:modified_cont}
\end{equation}
where $\tilde{D}_j = D_j(\eta r_0)^{-2}\varOmega^{-1}_0$ corresponds to the normalized diffusion coefficient for the $j$th dust-species.
Similarly, we modify the gas momentum equation to include the viscous term as follows:
\begin{align}
\partial_t \mathbf{v}_{\rm g} + \mathbf{v}_{\rm g} \cdot \nabla \mathbf{v}_{\rm g} &= 3\varOmega_0^2 x \mathbf{e}_x  -  2\varOmega_0 {\bf e}_z \times \mathbf{v}_{\rm g} \nonumber\\
 &+ {\bf F}_{\rm g} + {\bf a}_{\rm g} -  \frac{\nabla P}{\rho_{\rm g}}\, + \nu \nabla^2 {\bf v}_{\rm g} \,.
\end{align}
Therefore, the viscous term is included in the linearized gas momentum equation \eqref{Eq:lin_mom} by modifying\footnote{The gas continuity equation is not modified in this case.} $\tilde{\gamma}_{\rm g}$ as $\tilde{\gamma}_{{\rm g}} \equiv \gamma/\varOmega_0 - \tilde{{\bf k}} \cdot \tilde{\bf v}^0_{{\rm g}} + i\tilde{\nu} \tilde{k}^2$, where $\tilde{\nu} = \nu (\eta r_0)^{-2}\varOmega^{-1}_0$ the dimensionless gas viscosity. 
When including turbulence with Schmidt number $Sc = \nu/D = 1$ in a regime where $T_{\rm s} \ll 1$, the growth rate with turbulence agrees well with a simple correction, $\sigma_{\rm turb} \equiv \sigma_{\rm inviscid} - t^{-1}_{\rm visc}$, to the laminar growth rate, $\sigma_{\rm inviscid}$, with $t^{-1}_{\rm visc} = \nu k^2$.
This result agrees with \cite{Zhuravlev2020}, who also investigates other $Sc$ values not considered here.

\subsection{Stability for $\lambda \lesssim l_{\rm eddy}$ }\label{sec:subeddy}

For modes with $k > 2\pi/ l_{\rm eddy}$ the diffusion approximation cannot be applied.
Thus, to analyze the stability properties in this case, we follow \citetalias{Squire2018} in assigning a turbulent damping rate
\begin{align}
t^{-1}_{k} = \varOmega_0 \alpha^{1/3} (Hk/(2\pi))^{2/3}\, ,
\end{align} 
which is the inverse turnover time of eddies at scale $k$ in a Kolmogorov cascade.  We also consider the approximate turbulent growth condition of \citetalias{Squire2018}, that the non-turbulent growth rate must exceed the damping rate $t^{-1}_{k}$, that is modes with growth rate, $\sigma$, (in the absence of turbulence) satisfy $\sigma > t^{-1}_{k}.$

This comparison  is done using the analytical estimation of $\sigma$ found by \citetalias{Squire2018}. 
Assuming $z_0 = H$, for $\alpha \lesssim 10^{-3}$ species with Stokes numbers $T_{\rm s} \lesssim 0.005$ will have the double-resonant modes in a regime where the diffusion approximation can not be applied, while the rest of the resonant modes are at scales $\lambda > l_{\rm eddy}$. 
Thus, we adopt an estimate for the growth-rate of the form $\sigma \simeq (\epsilon T_{\rm s} k_x \eta r_0)^{1/3}$ with $k_x \sim k$, therefore our discussion differs from that of \citetalias{Squire2018} because they assumed $\sigma \simeq \sqrt{\epsilon}k_x/k$. 

Combining the wavelength and damping rate restrictions, the instability will grow when the double inequality,
\begin{align}
2\pi\eta r_0/l_{\rm eddy} < k_x \eta r_0 \lesssim (\eta r_0/H)^{2} \epsilon T_{\rm s}\alpha^{-1} \, , 
\end{align} 
is met. 
This condition implies that $\alpha > \left( \epsilon T_{\rm s} \eta r_0/(2\pi H)\right)^2$  leads to linear stability at scales $k_x \gtrsim 2\pi/l_{\rm eddy}$.
For instance, assuming $\epsilon = 0.1$ and $T_{\rm s}=10^{-2}$ we obtain a value of $\alpha \lesssim 10^{-11}$ for growth. Hence, at these scales we found a more stringent condition for stability in comparison with the results of Section\,\ref{sec:Turbulent_Background}.

\section{Numerical Test for Linear Growth}
\label{sec:Test}
\begin{figure}
	\includegraphics[]{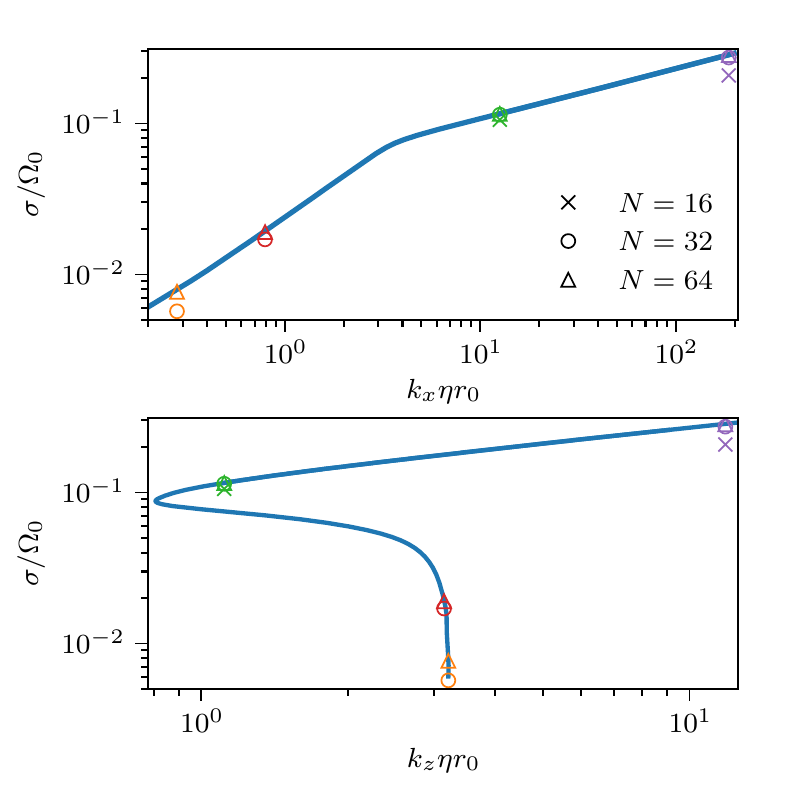}
	\caption{Numerical convergence test for the linear growth of the DSI with $\epsilon=10^{-2}$ and $T_{\rm s}=10^{-2}$. The solid blue curves correspond to the growth rate along the resonant modes.
		The top panel shows the growth rate as a function of $k_x\eta r_0$, while the bottom panel indicates the corresponding growth rates for the resonant $k_z$. 
		The cross, circle, and triangle symbols correspond to resolutions of 16, 32, or 64 cells per wavelength. 
		Accurate recovery of linear growth rates for large wavenumber modes requires only 16 cells per wavelength; for smaller wavenumbers ($\lambda \sim H$) at least $32$ cells per wavelength are necessary to capture growth rates. The higher resolution requirement is indicated by symbol clusters where crosses are not visible, as they are off the scale of the plot.}
	\label{fig:fig10}
\end{figure}
In this section we test the convergence of the growth-rate with the number of grid cells for simulations with FARGO3D.
We consider a dust-to-gas mass ratio of $\epsilon = 10^{-2}$ and Stokes number of $T_{\rm s}=10^{-2}$.
For a given pair of resonant wavenumbers $(k_x, k_z)$, each density and velocity perturbation is initialized with the eigenvector amplitudes as described in Appendix\,\ref{sec:linearized}.
These perturbations are added to the steady-state solutions.
The code setup up is equivalent to that adopted in Section\,\ref{sec:NonLinear} for the 2D simulations.
However, for the purpose of this test, vertical and radial size correspond to the wavelengths of the chosen mode.

In Fig.\,\ref{fig:fig10} we show the results obtained for three different resolutions. 
Modes were well-resolved with either $N = 16$ or $N = 32$ cells per wavelength.  Higher resolution is required for longer wavelengths (small $k_x$) presumably because slower growth rates are more  numerically challenging.

\bsp	
\label{lastpage}
\end{document}